\documentclass[twocolumn]{emulateapj}
\usepackage{color}
\usepackage{url}
\usepackage[usenames,dvipsnames,svgnames,table]{xcolor}
\usepackage[colorlinks=true,
           linkcolor=blue,
           urlcolor=blue,%
           citecolor=blue]{hyperref}

\begin{document}

\title{FLARE GENERATED SHOCK WAVE PROPAGATION THROUGH SOLAR CORONAL ARCADE LOOPS AND ASSOCIATED TYPE II RADIO BURST}
\author{PANKAJ KUMAR\altaffilmark{1,2}, D.E. INNES\altaffilmark{2}, KYUNG-SUK CHO\altaffilmark{1,3}}
\affil{$^1$Korea Astronomy and Space Science Institute (KASI), Daejeon, 305-348, Republic of Korea}
\affil{$^2$Max-Planck Institut f\"{u}r Sonnensystemforschung, D-37077 G\"{o}ttingen, Germany}
\affil{$^3$University of Science and Technology, Daejeon 305-348, Korea}
\email{pankaj@kasi.re.kr}

\begin{abstract}
This paper presents multiwavelength observations of a flare-generated type II radio burst. The kinematics of the  shock derived from the type II closely match  a fast EUV wave seen propagating through coronal arcade loops. The EUV wave was closely associated with an impulsive M1.0 flare without a related coronal mass ejection, and was triggered at one of the footpoints of the arcade loops in active region NOAA 12035. It was initially observed in the AIA 335~\AA\ images with a speed of $\sim$800~km~s$^{-1}$ and accelerated to $\sim$1490~km~s$^{-1}$ after passing through the arcade loops. A fan-spine magnetic topology was revealed at the flare site. A small, confined filament eruption ($\sim$340~km~s$^{-1}$) was also observed moving in the opposite direction to the EUV wave. We suggest that breakout reconnection in the fan-spine topology triggered the flare and associated  EUV wave that propagated as a fast shock through the arcade loops.  

\end{abstract}

\section{INTRODUCTION}
Type II radio bursts are considered to be an indirect signature of shocks \citep{nelson1985,vrsnak2008}. Most of the coronal type II radio bursts are excited by  coronal mass ejection (CME) pistons \citep{thompson2000,cliver1999,gopal2001}. Another source of type II radio bursts are shocks generated by blast waves created by a pressure increase at the flare site \citep{uchida1974,hudson2003,hudson2004}.
The flare impulsive phase and CME acceleration phase often occur simultaneously \citep{zhang2001}, therefore, it is difficult to confirm the exact driver of coronal shocks in terms of flare blast or CME piston. An argument in favour of the CME driver is that confined flares (even  X-class flares) do not produce coronal shocks. On the other hand, shocks are only seen when the associated  CME is accompanied by a flare \citep{cliver2004,gopal2005}.

In support of the flare-driven shock, some studies using SOHO/EIT and LASCO coronagraph images suggest the formation of a type II radio burst during the flare impulsive phase when the CME acceleration phase starts 10-20 min later,  or the CME is too slow to drive a shock wave ($<$500 km s$^{-1}$)  \citep{mag2008,mag2010,mag2012}. The shock waves in such events were associated with impulsive/short duration flare events with rise times less than 4~min. 
These studies have used low resolution SOHO/EIT images (12 min cadence) and it was therefore not possible to track the low coronal eruptions (plasmoid/jets) related  to short duration impulsive flares and their associated shock waves ($\sim$1000 km $^{-1}$) in extreme ultraviolet (EUV) images.
Now data from the  {\it Atmospheric Image Assembly} (AIA; \citealt{lemen2012}) on board the {\it Solar Dynamics Observatory} (SDO; \citealt{pesnell2012}) provides an opportunity to study fast EUV waves and associated plasma ejections in the low corona and it  is therefore possible to investigate the wave driver and kinematics in more detail.

\citet{kumar2013blast} provided evidence of a blast wave ($>$1000 km s$^{-1}$) propagating alongside and ahead of an erupting plasmoid that produced a metric type II radio burst.
Recently, \citet{kumar2015w} reported a clear observation of a fast-mode wave propagating along arcade loops and its partial reflection from the other footpoint. They also observed a second fast EUV wavefront propagating perpendicular to the arcade loops which triggered a metric type II radio burst. The wave was excited by an impulsive C-class flare (without a CME) at one of the footpoints of the arcade loops.

Solar eruptions are generally associated with EUV waves. These waves were discovered by  SOHO/EIT  \citep{thompson1998}. There has been a long-lasting debate about their nature (true wave or pseudo wave) and driver (in terms of CME or flare) (\citealt{vrsnak2008,veronig2010, warmuth2010,warmuth2011,pats2012,liu2014,warmuth2015} and references cited therein).
At present, high resolution observations from  SDO/AIA suggest the existence of two wavefronts (i.e., fast and slow) \citep{chen2011,kumar2013b,kumar2013blob}. The fast wavefront speed (of the order of $\sim$1000 km s$^{-1}$) is about three times larger than the slow one as predicted in the numerical simulation by \citet{chen2002}. The fast wavefront is either a  true fast-mode wave or an magnetohydrodynamic (MHD)  shock ahead of the slow wavefront. The slow wavefront is basically expanding CME loops interpreted as a pseudo wave (non-wave component). The fast EUV wavefront triggers transverse oscillations in coronal loops 
\citep{wills1999,asc1999,sch2002,asc2002,kumar2013b} 
or filament channels when passing through these structures \citep{asai2012,shen2014} and shows reflection, refraction and transmission when it interacts with neighboring active regions \citep{kumar2013blob} or coronal holes \citep{gopal2009,olmedo2012}. On the other hand, the slow wavefront generally stops at the boundary of an active region or at magnetic separatrices \citep{delanee1999}, therefore, supporting the non-wave interpretation.

In this paper, we study a type II radio burst associated with a fast EUV wave observed during an impulsive C-class flare (without a CME) on 16 April 2014. The EUV wave propagates through arcade loops and triggers transverse oscillations in the loop system. 
We focus on the wave characteristics, its propagation, and excitation. In section 2, we present the observation and results. In the last section, we discuss the results and draw some conclusions. 

\section{OBSERVATIONS AND RESULTS}

\subsection{Radio spectrum}
\begin{figure*}
\centering{
\includegraphics[width=12cm]{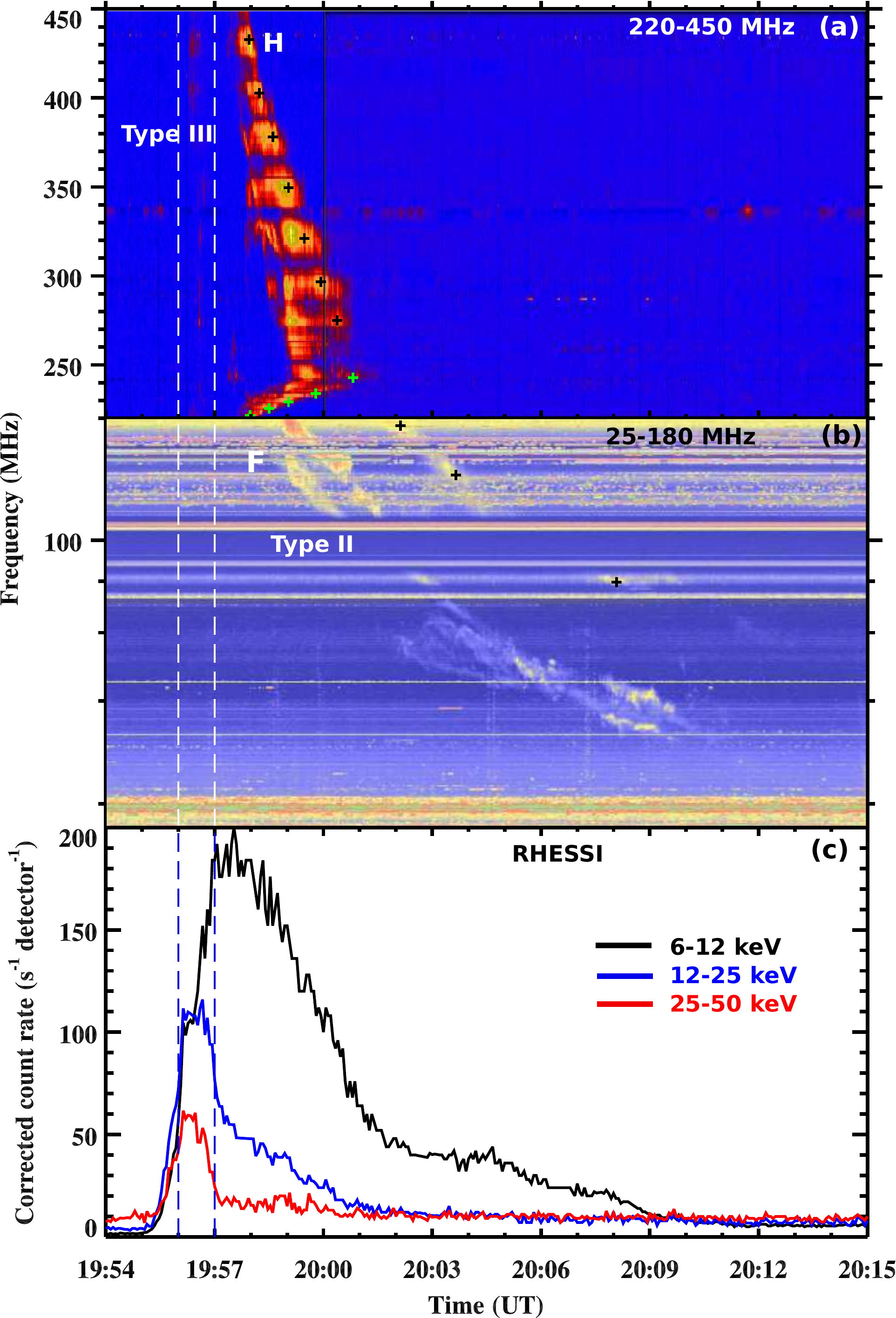}
}
\caption{(a,b) Dynamic radio spectrum from e-Callisto (Roswell station, 220-450 MHz) and RSTN (Sagamore hill station, 25-180 MHz). F and H denote the fundamental and second harmonic of the type II burst. The black pluses mark the frequencies used for obtaining the type II exciter speed. The green pluses mark the frequencies for calculating the speed of the positive drift feature. (c) RHESSI X-ray flux profiles in 6-12, 12-25, and 25-50 keV channels. Two vertical dotted lines represent the onset time of nonthermal particle acceleration in the radio (type III burst) and hard X-ray (12-50 keV) channels.}
\label{spectrum}
\end{figure*}

The type II burst was seen in the dynamic radio spectra from e-Callisto (Roswell station, 220-450 MHz) and RSTN (Sagamore hill station, 25-180 MHz) from about 19:57:40~UT to 20:10~UT (Figure~\ref{spectrum}(a,b)). The composite spectrum shows both the fundamental and harmonic of the type II burst. 
The fundamental shows band splitting probably caused by emission from the pre- and post-shock plasma  \citep{vrsnak2001,vrsnak2002,cho2007}. The splitting is typical of that seen in other CME-associated type IIs and corresponds to a compression ratio of 1.6 around 120~MHz and 1.4 around 40~MHz. 

\begin{figure}
\centering{
\includegraphics[width=9cm]{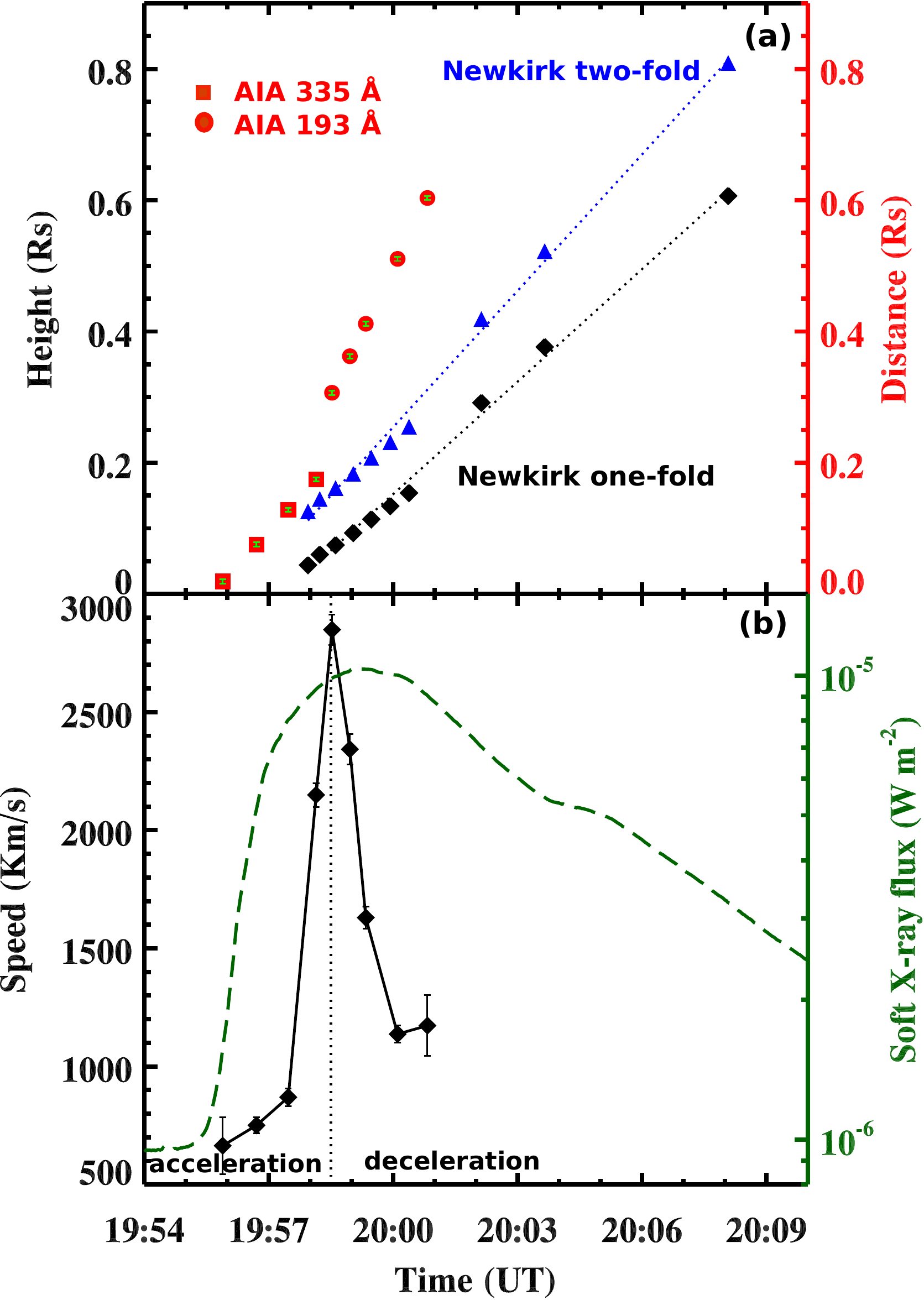}
}
\caption{(a) Time-height plots (R$_{\odot}$, above the solar surface) estimated from the Newkirk one-(diamond) and two-fold (triangle) density models using the second harmonic of the type II radio burst. The estimated shock speed from the linear-fit is $\sim$670 and $\sim$810 km s$^{-1}$, respectively, from the Newkirk one- and two-fold density models. The distance of the EUV wave from the flare center (Figure \ref{aia_stack}) using AIA 335 (filled diamond) and 193 (filled circle) images is over-plotted in red. (b) EUV wave speed profile estimated from the time-distance values (red) taken from panel (a). GOES soft X-ray flux in 1-8 \AA~ channel is also included (dark green).} 
\label{h-t}
\end{figure}

 We used the Newkirk one and two-fold density model \citep{newkirk1961} to estimate the height of the type II exciter by selecting a few data points (marked by the + symbol) from the second harmonic of the type II burst. The time-height plot is shown in Figure \ref{h-t}(a). The average speed (from the linear fit) of the type II emission source is $\sim$670 and $\sim$810 km s$^{-1}$ for the one- and two-fold models, respectively. 

An unusual feature of the radio spectrum is a positive drifting feature from  $\sim$220 MHz to $\sim$250 MHz starting simultaneously with the type II, which implies that  there were  sunward propagating energetic electrons at the same time as the outward moving shock creating the type II. 
To estimate the exciter speed, we selected the data points marked with a green + sign and converted the frequencies into the emission heights in the corona using the Newkirk density models (considering fundamental emission). The downward moving source speed (by a linear fit to the emission heights) is $\sim$110 and $\sim$130 km s$^{-1}$ from the Newkirk one-fold and two-fold density models, respectively.

There was also a type III radio burst at high frequencies (220-450 MHz) but not below 200~MHz suggesting acceleration of sub-relativistic electrons that  did not escape into interplanetary space.  
 
 To find the source of the type II and other features of the radio spectrum, we investigated EUV images and magnetic field data from  AIA and the {\it Heliospheric and Magnetic
Imager} (HMI; \citealt{schou2012}) on SDO,  images and spectra from Hinode's XRT (X-Ray Telescope; \citealt{golub2007}) and EIS (EUV Imaging Spectrometer; \citealt{culhane2007}), hard X-ray images from RHESSI (Reuven Ramaty High Energy Solar Spectroscopic Imager, \citealt{lin2002}), and SOHO/LASCO \citep{brueckner1995,yashiro2004} and STEREO-B COR-1 \citep{wuelser2004,howard2008} coronagraph images. 
 
 At the time of the type II, there was an M1.0 flare in  active region NOAA 12035, when it was located in the southeast (S19E12) with a $\beta\gamma$ magnetic configuration. The M1.0 flare was an impulsive short duration flare which started at 19:54 UT, peaked at 19:59, and ended at 20:04 UT.
Figure \ref{spectrum}(c) shows the RHESSI X-ray flux in the 6-12 keV, 12-25 keV, and 25-50 keV channels. The type III radio burst correlates with the hard X-ray flux in 12-25 and 25-50 keV channels, suggesting the emission from non-thermal electrons. The 6-12 keV flux peaked slightly later ($\sim$30 sec), which indicates the contribution from the thermal emission. 

 A  careful check of the coronagraph images revealed  that there was no CME associated with this flare; however the EUV images showed a fast, outward propagating wave and the possible source of the positive drifting feature.
This study utilizes AIA 94~\AA\ (\ion{Fe}{10}, \ion{Fe}{18}, $T\approx$1 MK, $T\approx$6.3 MK), 
131~\AA\ (\ion{Fe}{8}, \ion{Fe}{21}, \ion{Fe}{23}, i.e., 0.4, 10, 16 MK), 
171~\AA\ (\ion{Fe}{9}, $T\approx$0.7 MK),  
 193~\AA\ (\ion{Fe}{12}, \ion{Fe}{24}, $T\approx$1.5, 20~MK), 
304~\AA\ (\ion{He}{2}, T$\approx$0.05 MK),
335~\AA\ (\ion{Fe}{16}, T$\approx$2.5~MK),
and 1600~\AA\ (\ion{C}{4} + continuum, $T\approx$0.1 MK) images. 

\begin{figure*}
\centering{
\includegraphics[width=6.8cm]{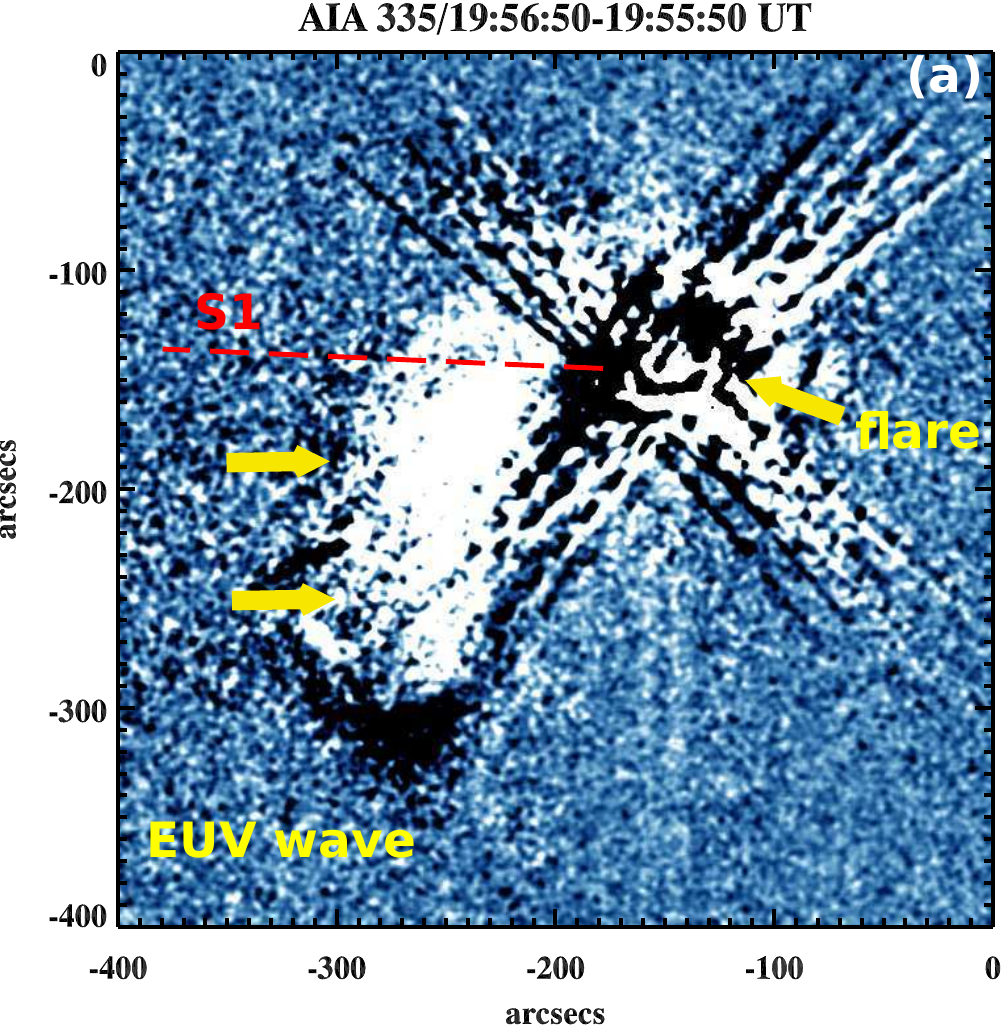}
\includegraphics[width=6.5cm]{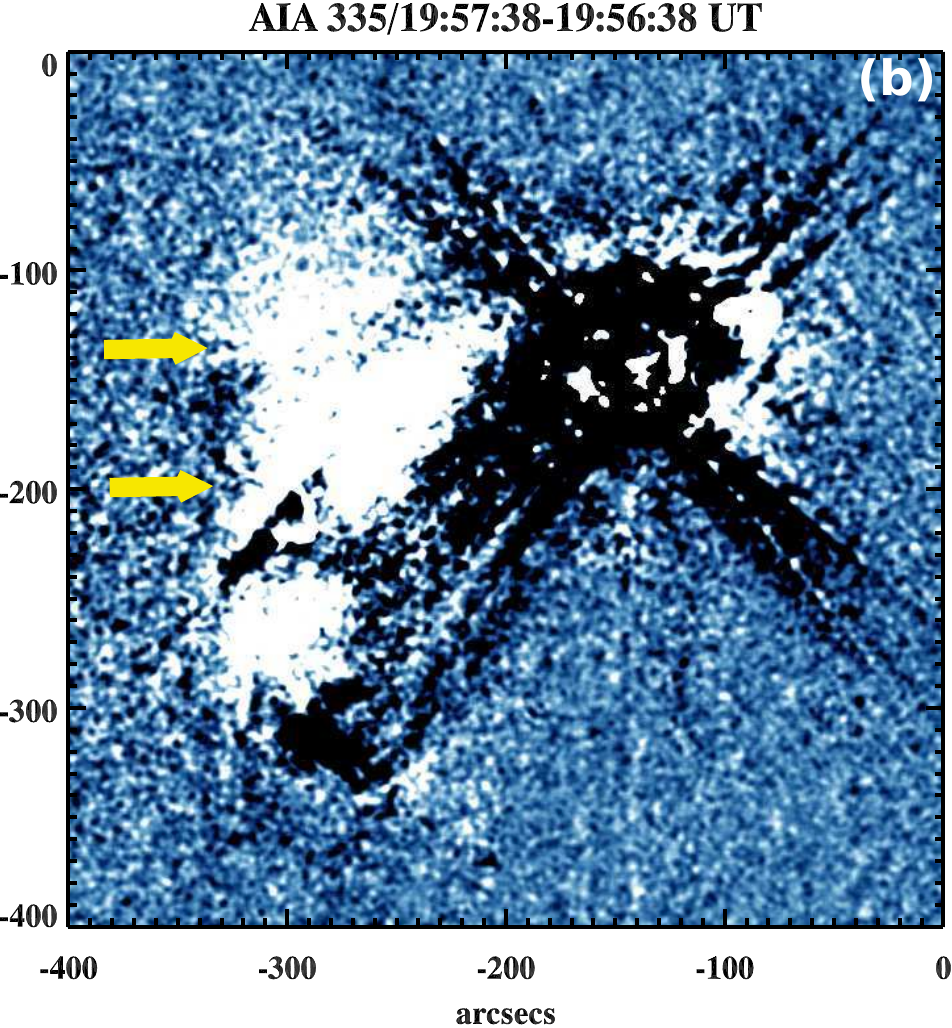}

\includegraphics[width=6.8cm]{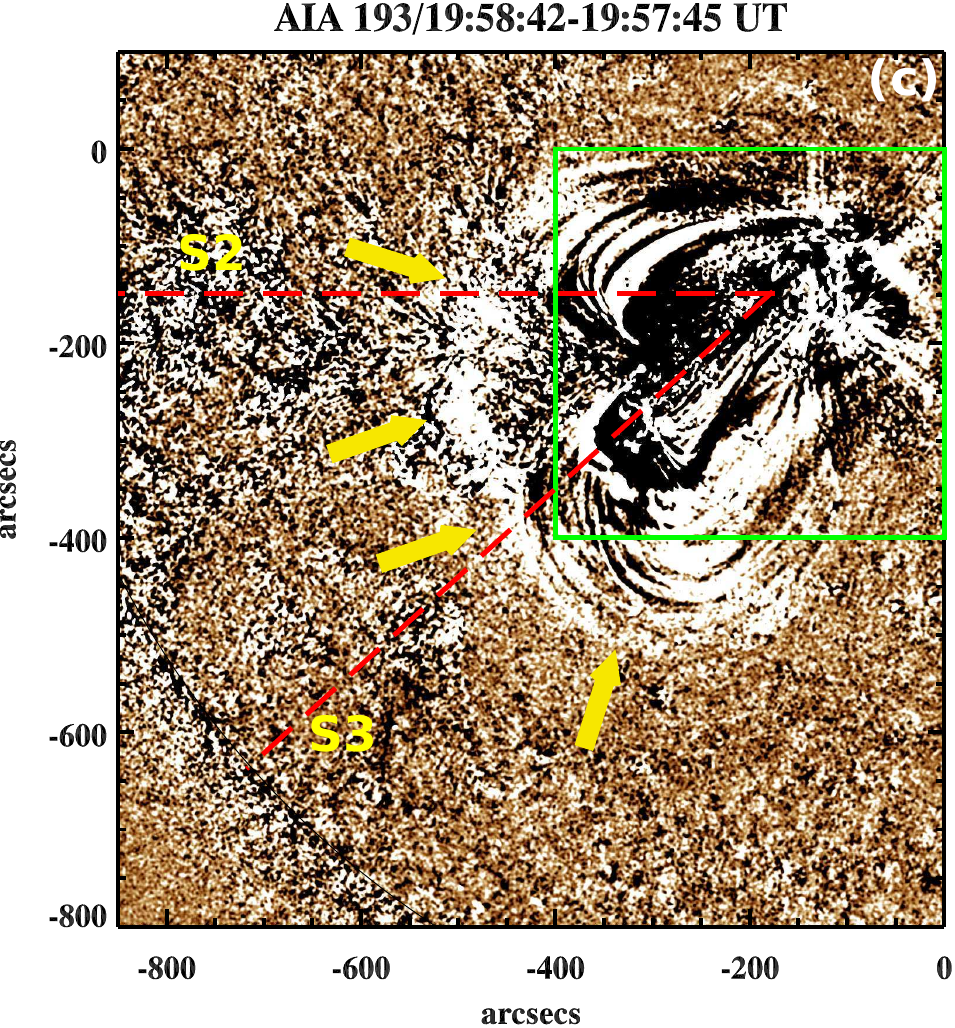}
\includegraphics[width=6.5cm]{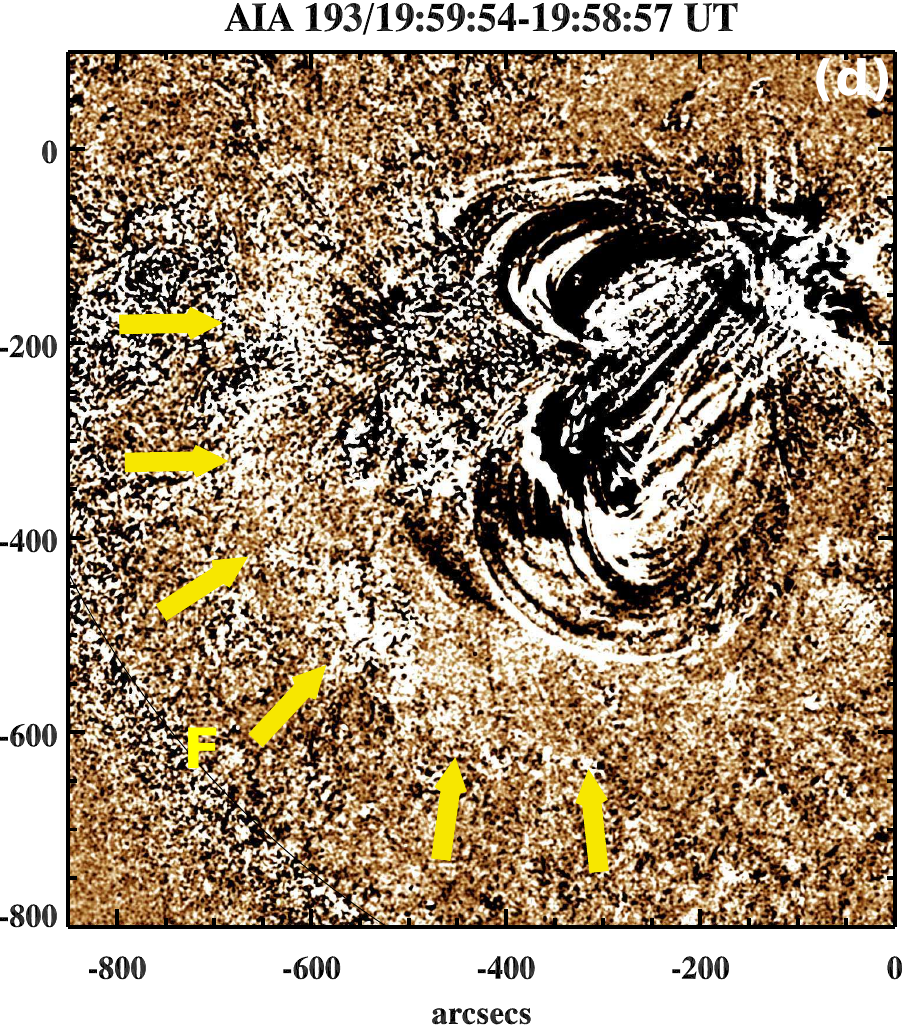}

\includegraphics[width=6.8cm]{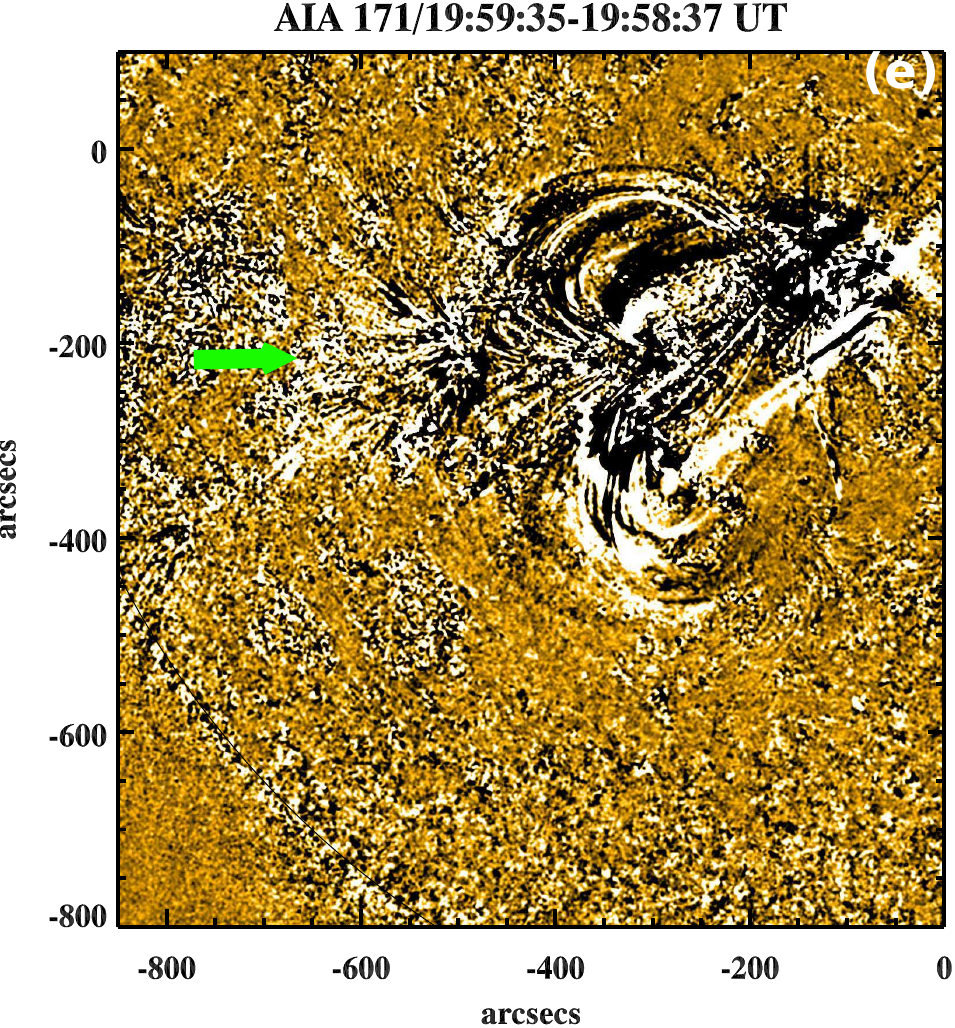}
\includegraphics[width=6.5cm]{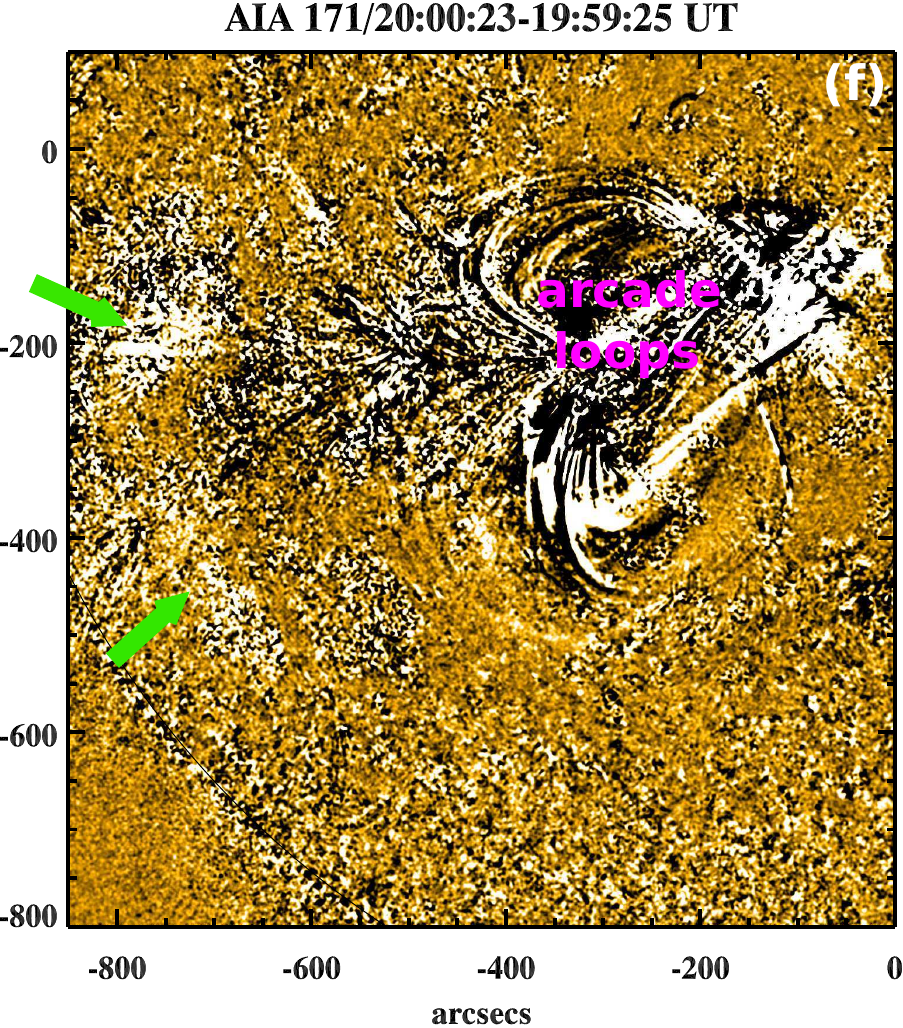}

}
\caption{AIA 335, 193 and 171 \AA~ running difference images showing the propagating EUV disturbance (marked by arrows) left to the flare site. The green box in panel (c) represents the size of the top panels (a,b). (An animation of this figure is available)}
\label{aia-rd}
\end{figure*}
\begin{figure}
\centering{
\includegraphics[width=9cm]{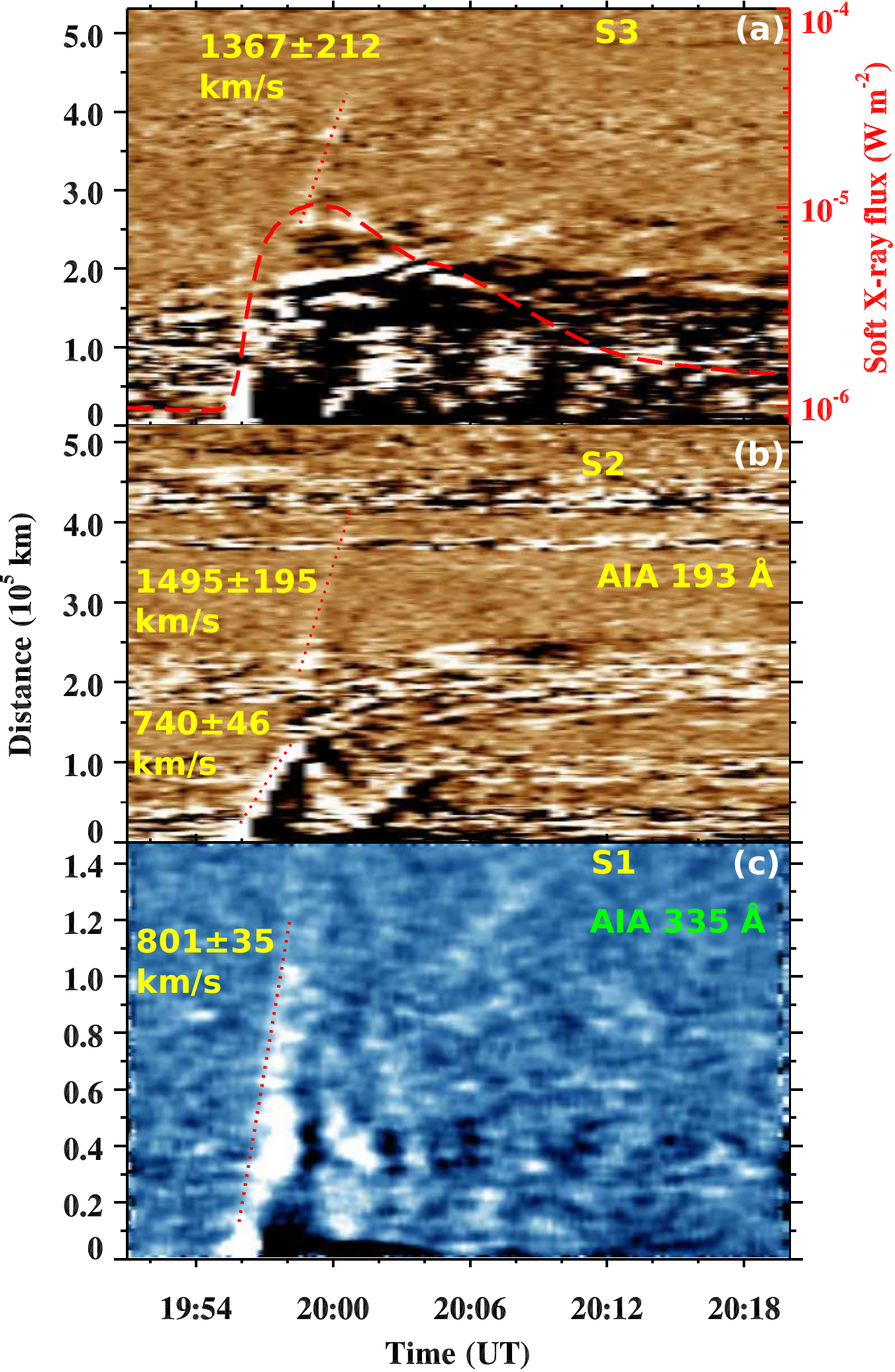}
}
\caption{Time-distance intensity (running difference) plots along the slices S1 (AIA 335 \AA), S2, and S3 (AIA 193 \AA). The red dashed curve in panel (a) shows the GOES soft X-ray flux profile in 1-8 \AA~ channel.}
\label{aia_stack}
\end{figure}

\begin{figure*}
\centering{
\includegraphics[width=8.0cm]{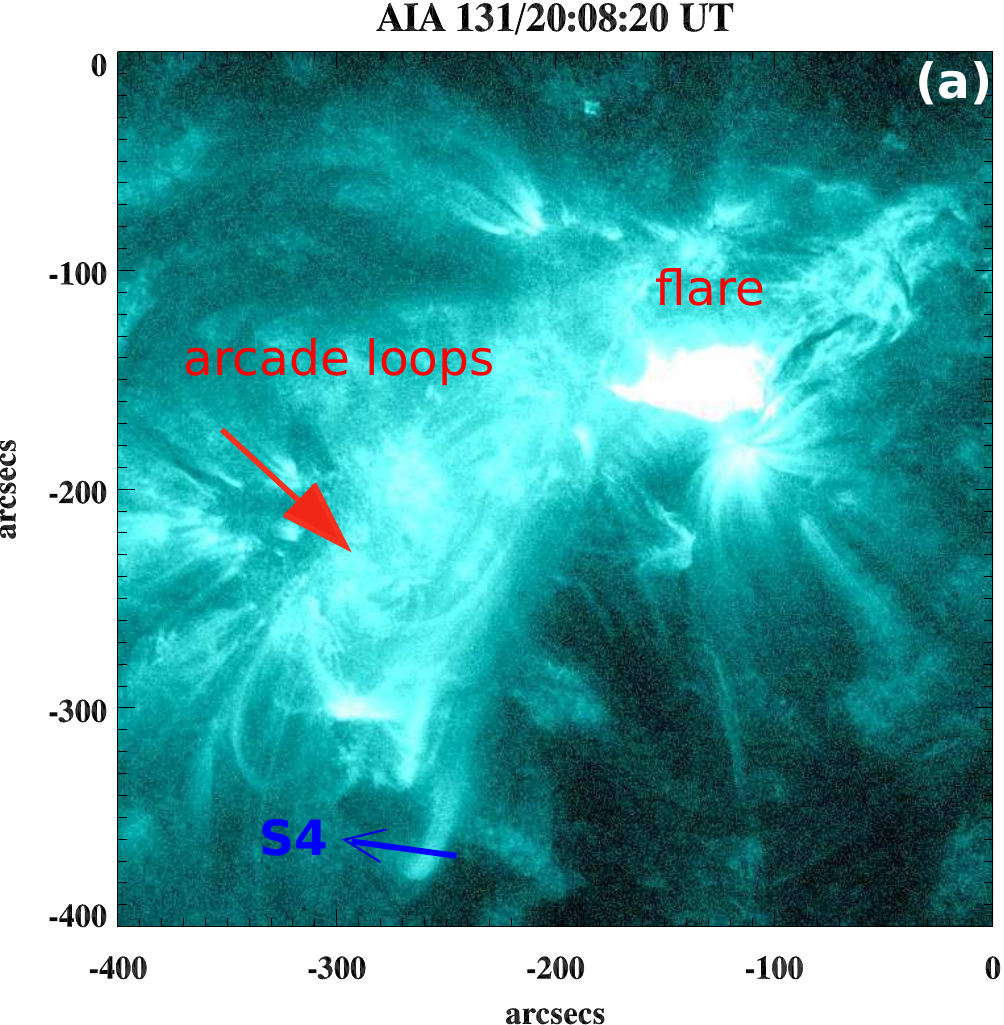}
\includegraphics[width=8.0cm]{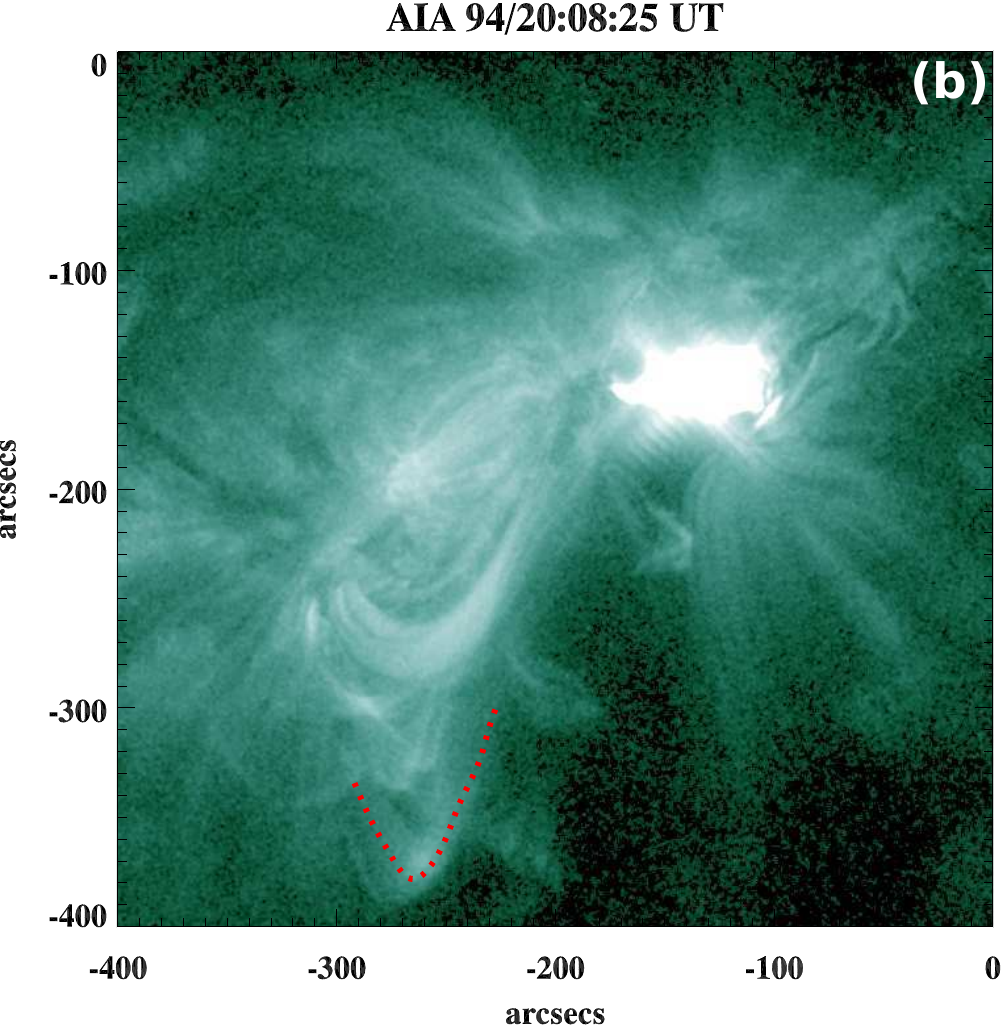}

\includegraphics[width=10cm]{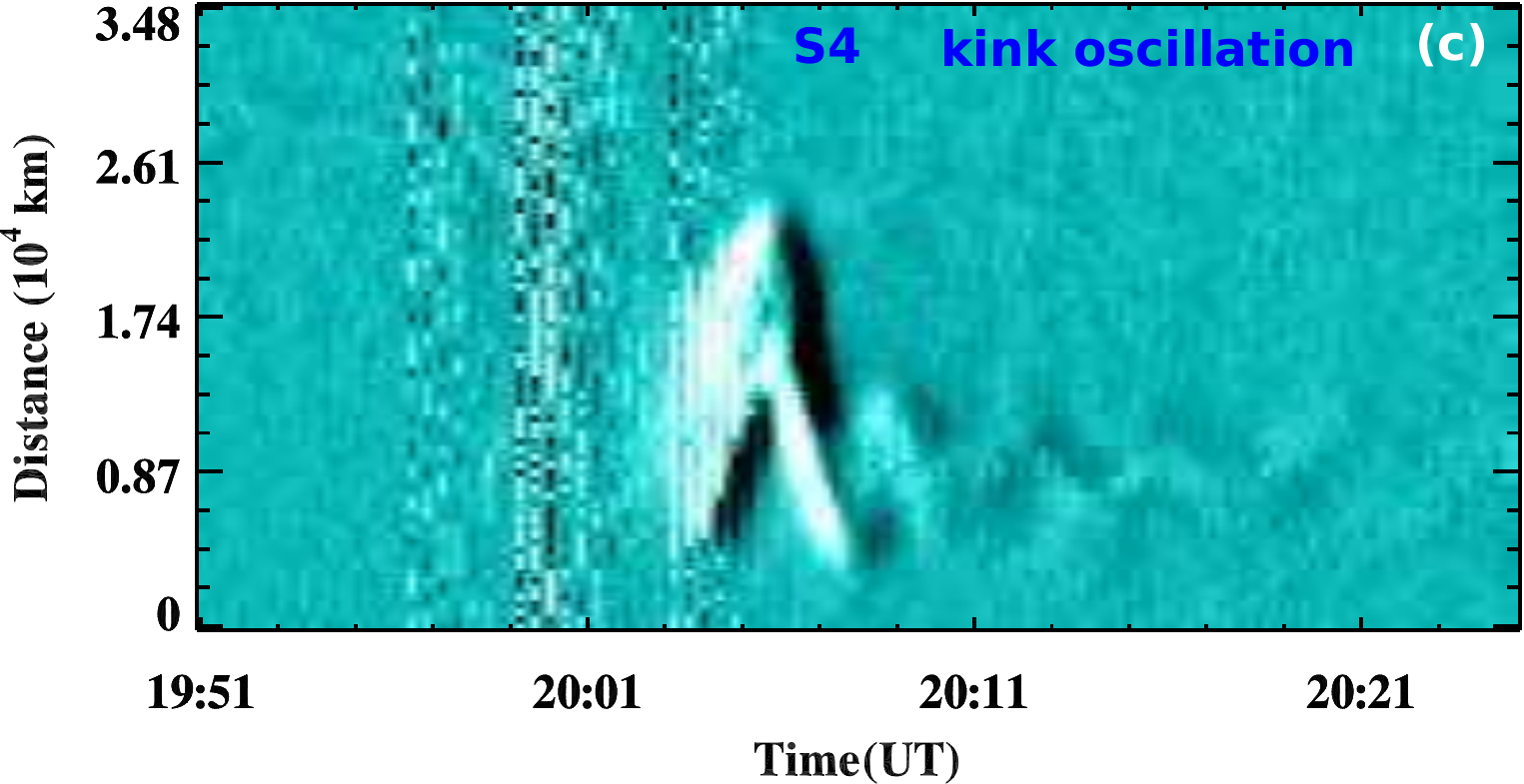}

\includegraphics[width=10cm]{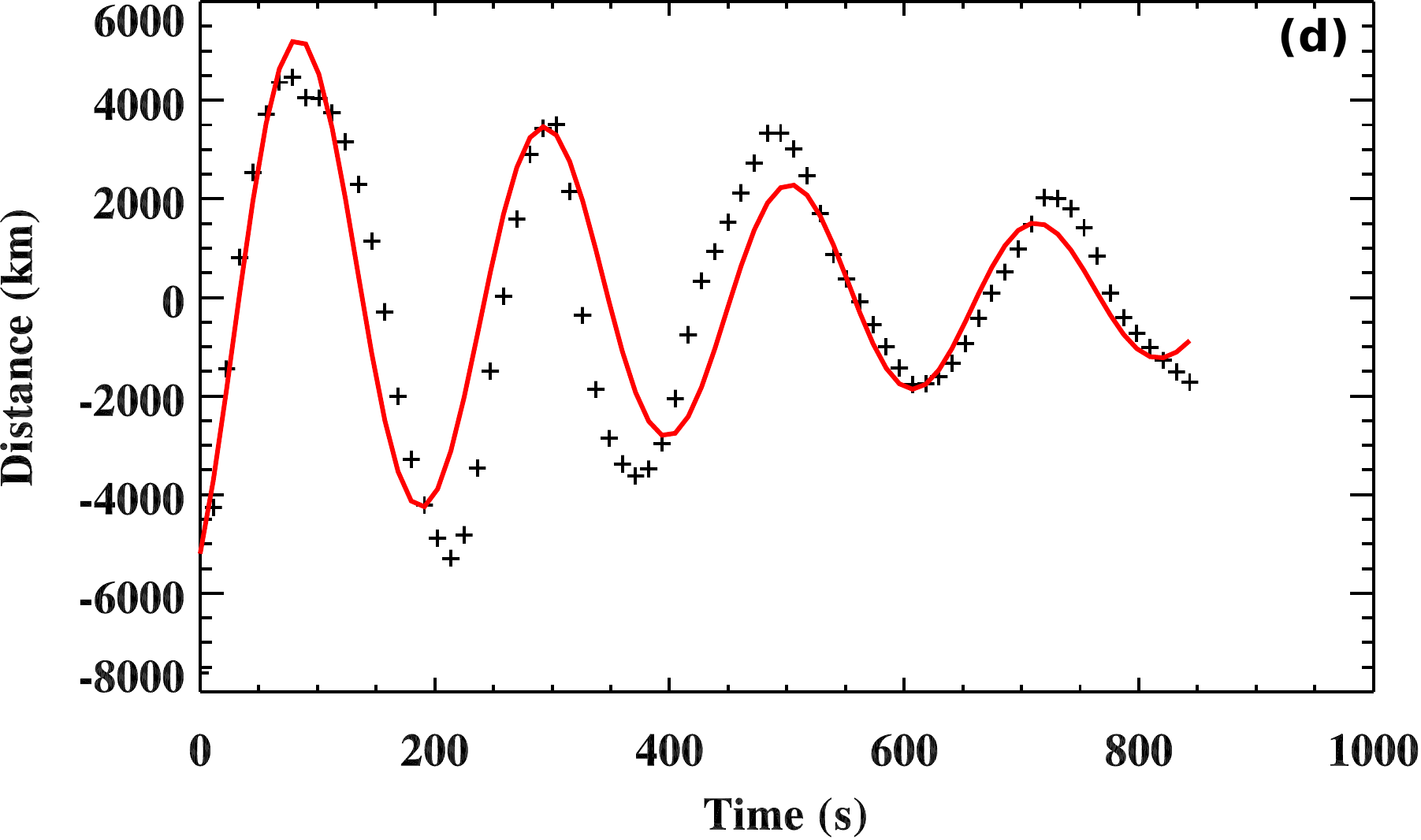}
}
\caption{(a,b) AIA 131 and 94 \AA~ images showing the flare and arcade loops. (c) AIA 131 \AA~ intensity (running difference) plot along the slice S4 (marked in panel(a)) showing the kink oscillation of the arcade loops. (d) Decaying sine function fit (red color) to the kink oscillating loop. The period of oscillation is $\sim$210 s. The start time is $\sim$20:04 UT. (An animation of this figure is available)}
\label{osc}
\end{figure*}
\begin{figure*}
\centering{
\includegraphics[width=16cm]{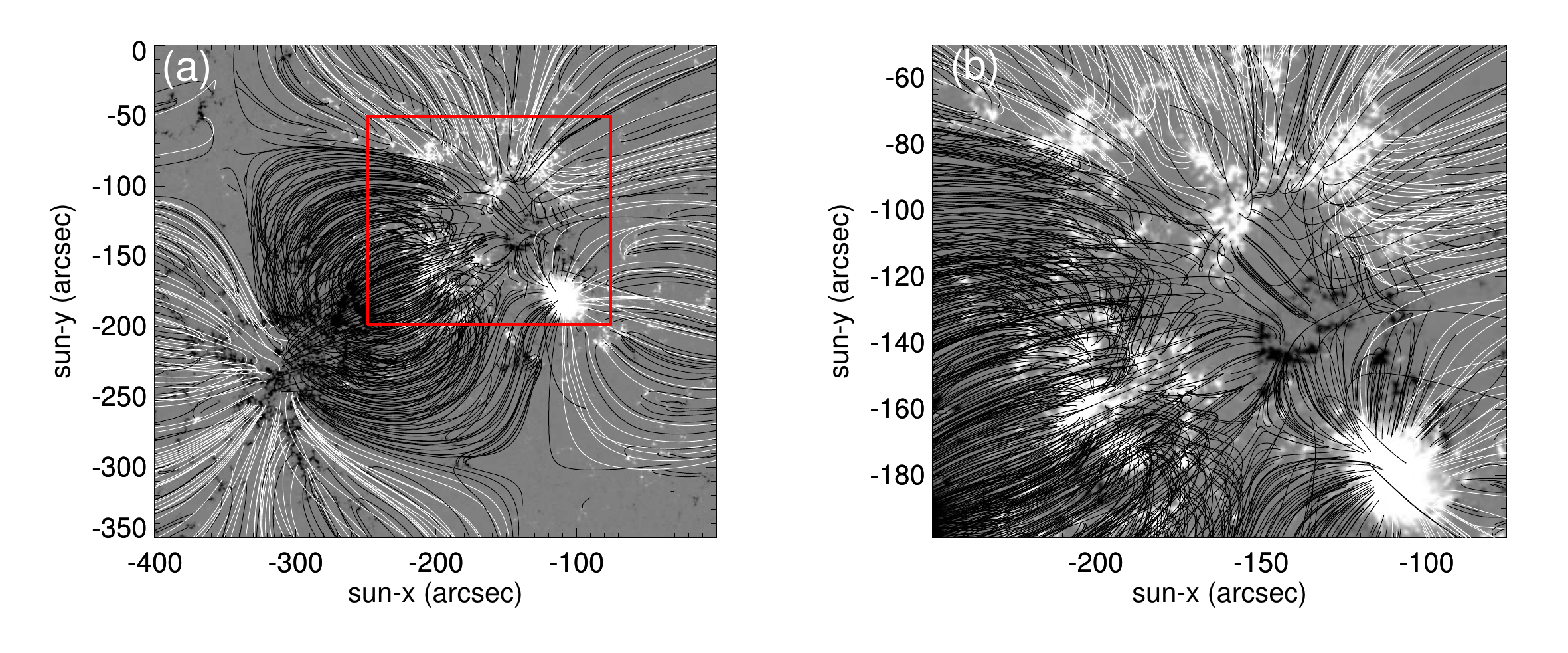}

\includegraphics[width=7.0cm]{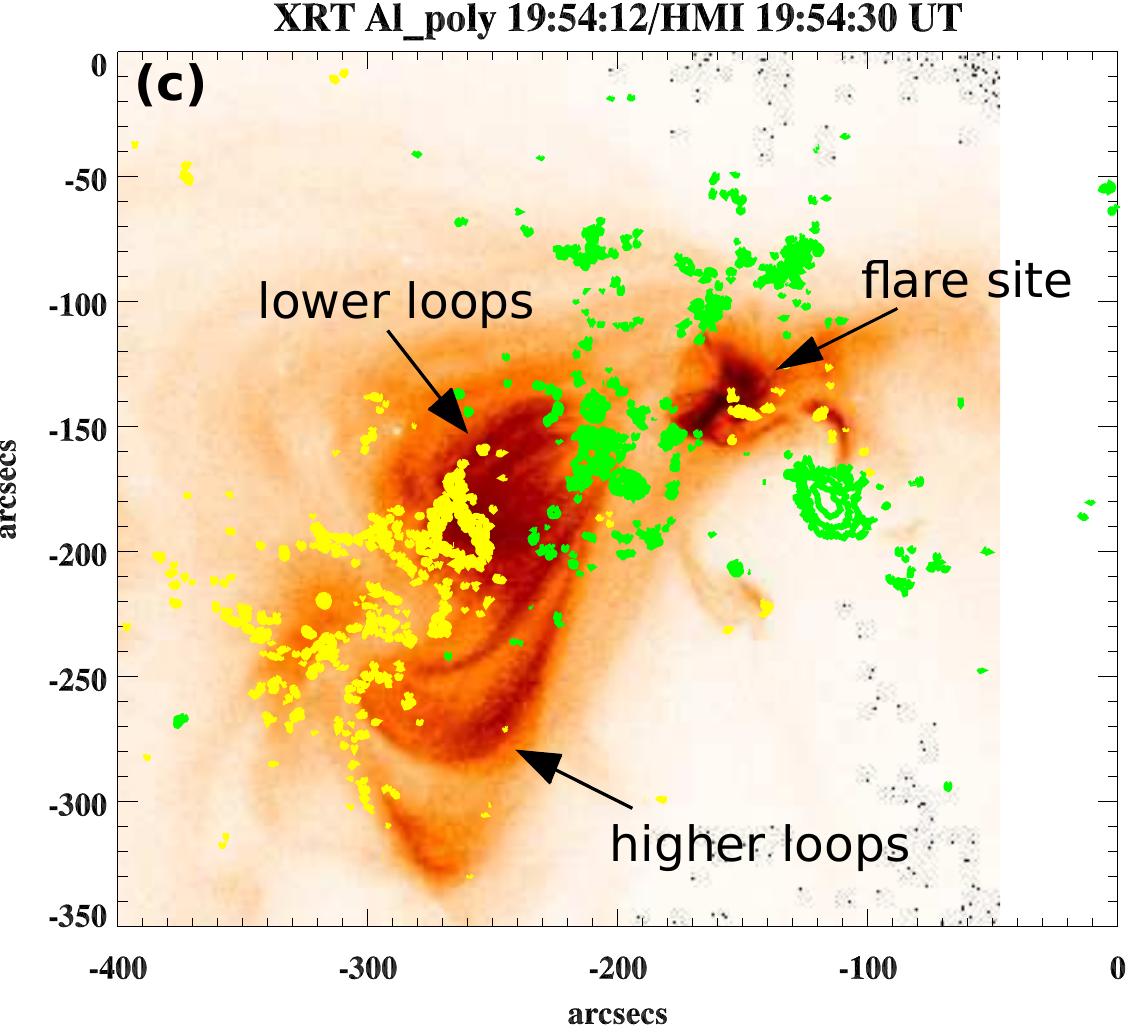}
\includegraphics[width=7.0cm]{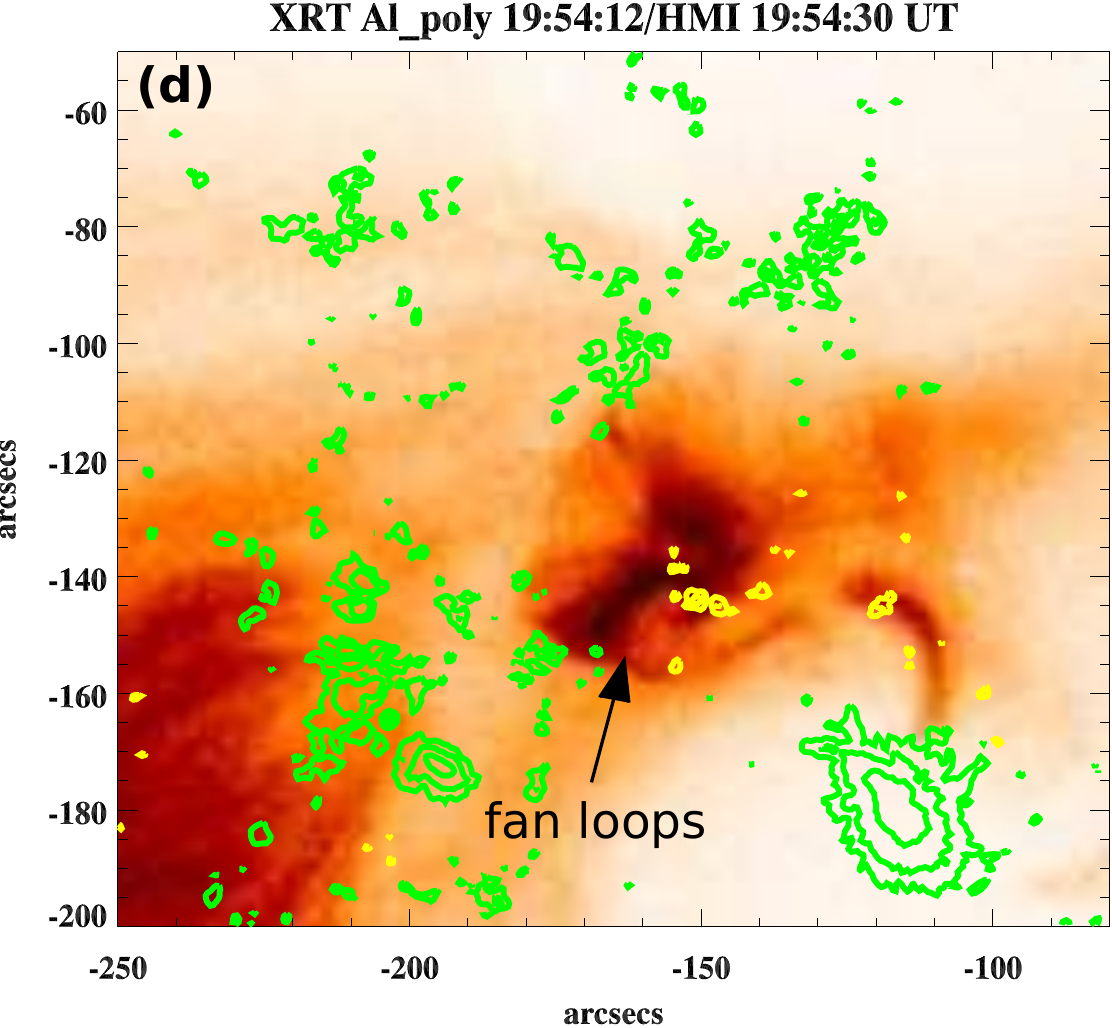}

\includegraphics[width=7.0cm]{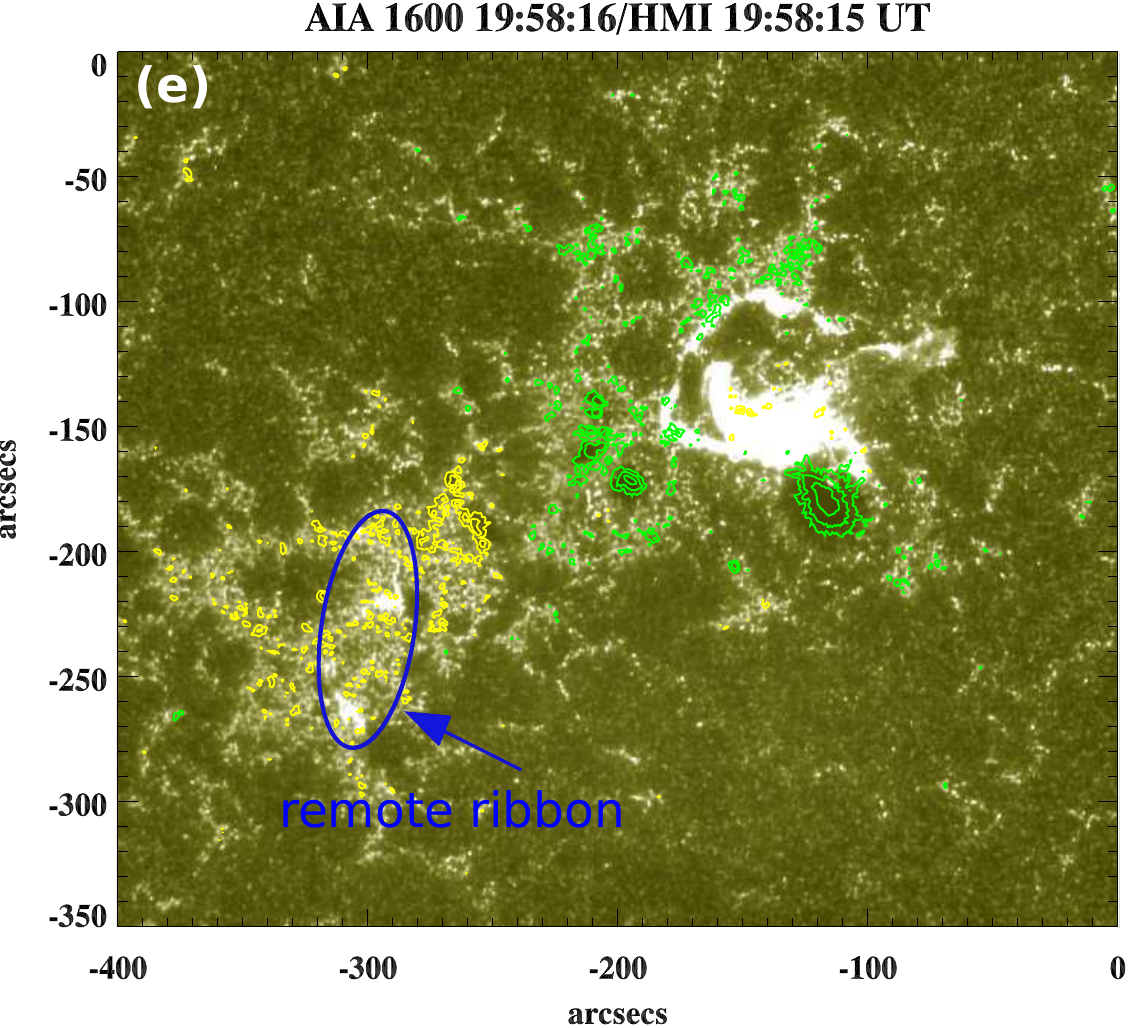}
\includegraphics[width=7.0cm]{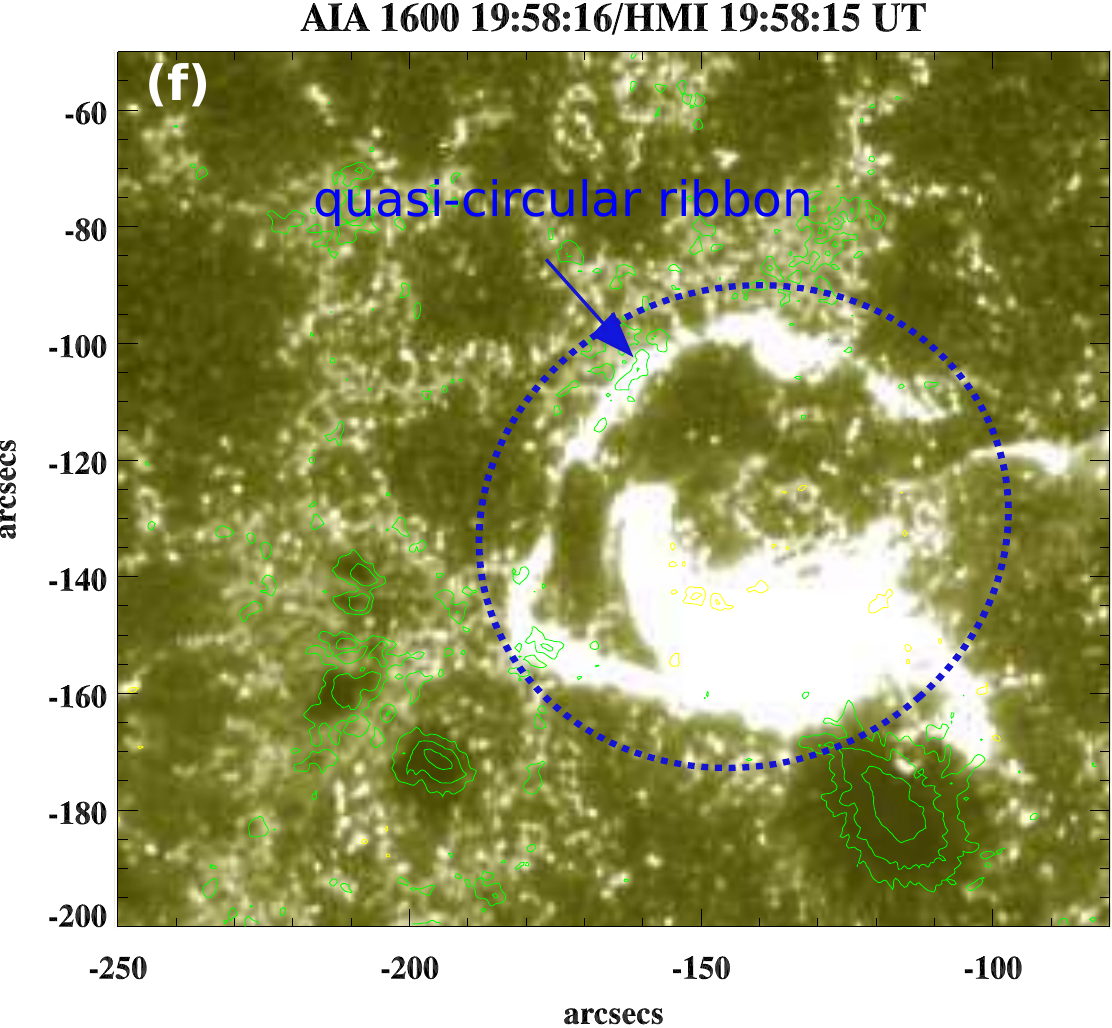}
}
\caption{(a,b) Potential field extrapolation of the active region (at $\sim$19:45 UT) showing fan loops at the flare site. The closed and open field lines are shown by the black and white color, respectively. The fast-mode wave propagated through the closed loops (eastward) and triggered kink oscillation of the arcade loops. The red box in panel (a) represents the size of panel (b). (c,d) Hinode/XRT image of the active region overlaid by HMI magnetogram contours of positive (green) and negative (yellow) polarities. (e,f) AIA 1600 \AA~ image showing the quasi-circular flare ribbon. The contour levels are $\pm$500, $\pm$1000, and $\pm$1500 Gauss.}
\label{mf}
\end{figure*}
\begin{figure*}
\centering{
\includegraphics[width=15cm]{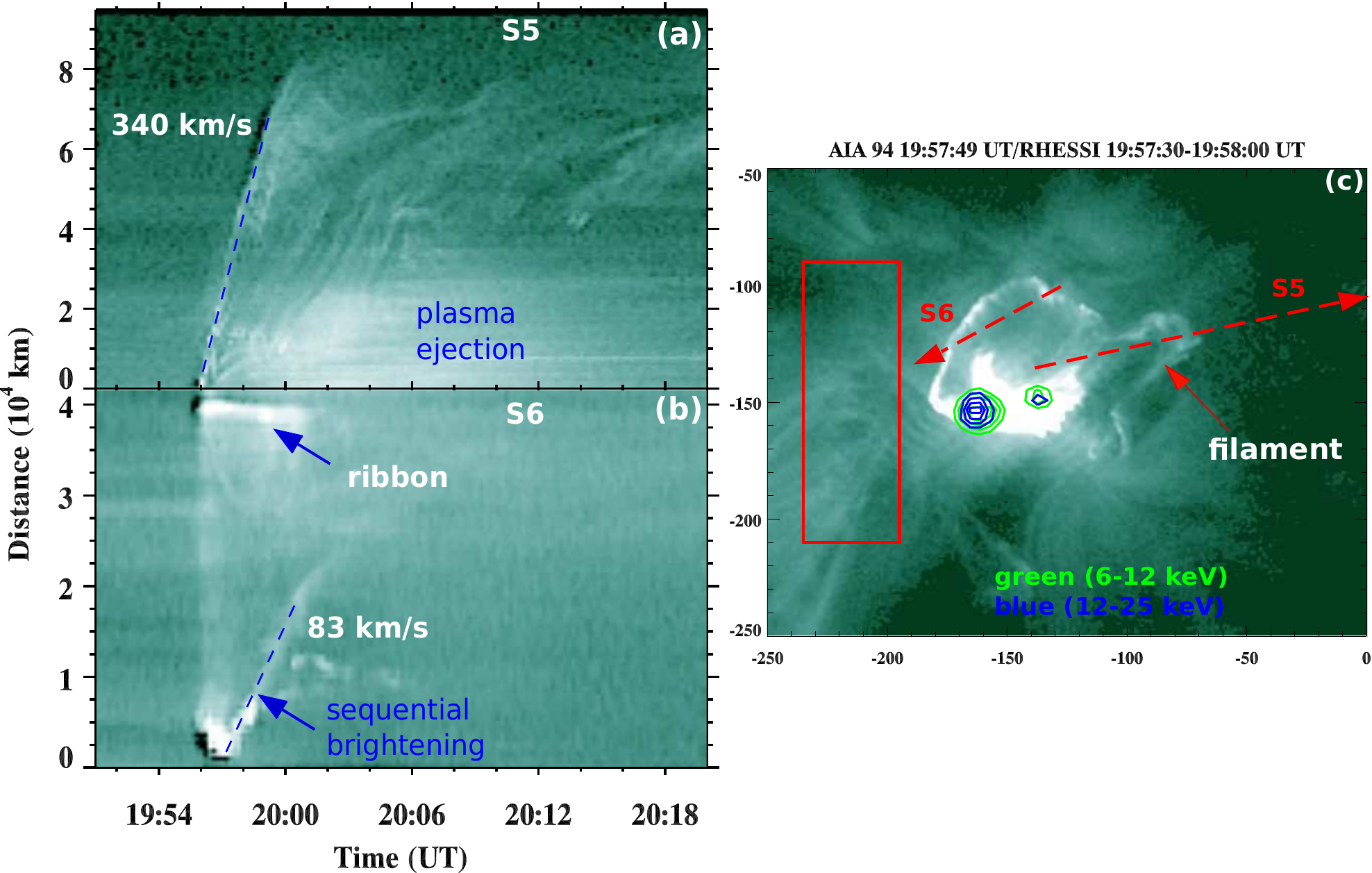}

\includegraphics[width=5.3cm]{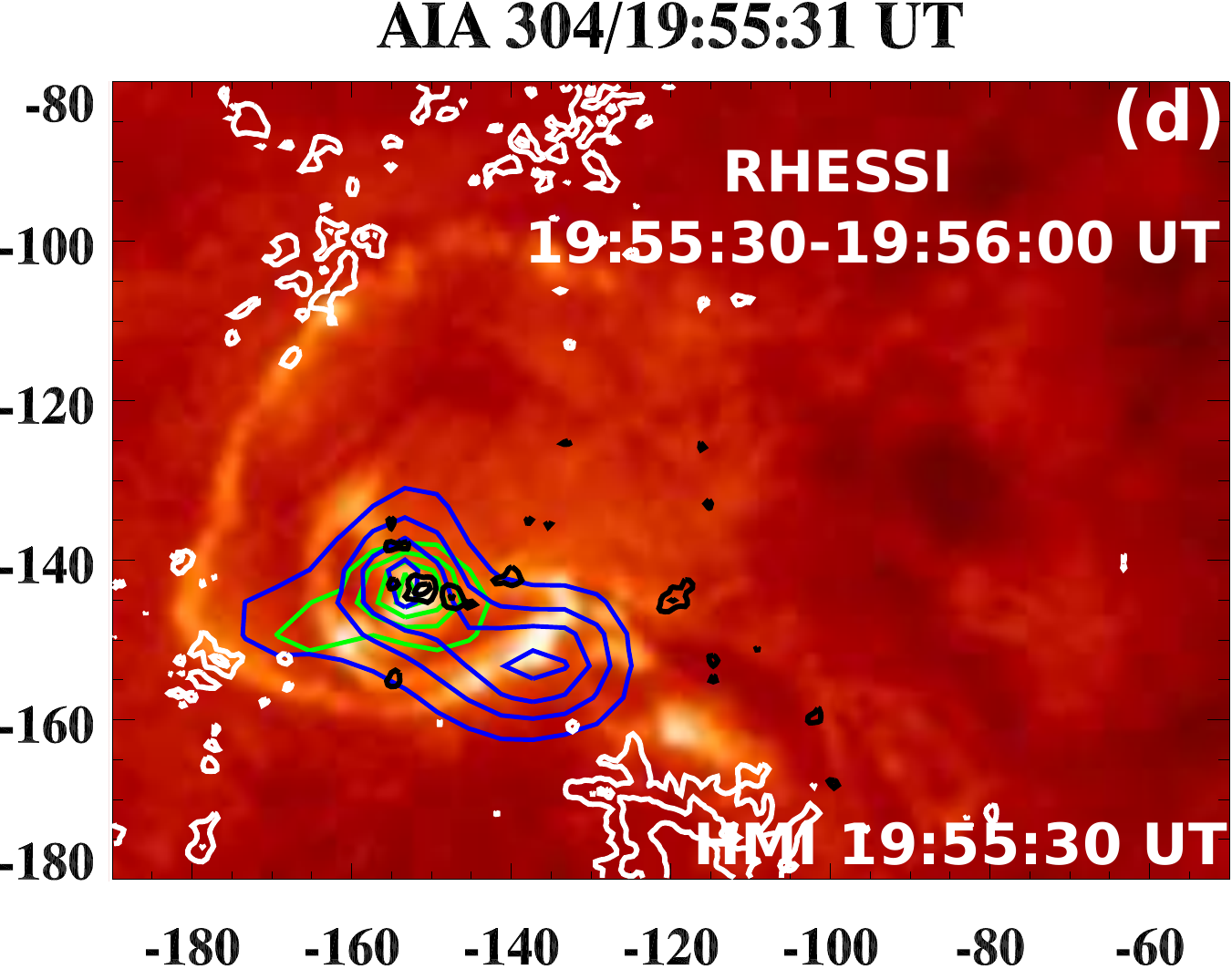}
\includegraphics[width=5.3cm]{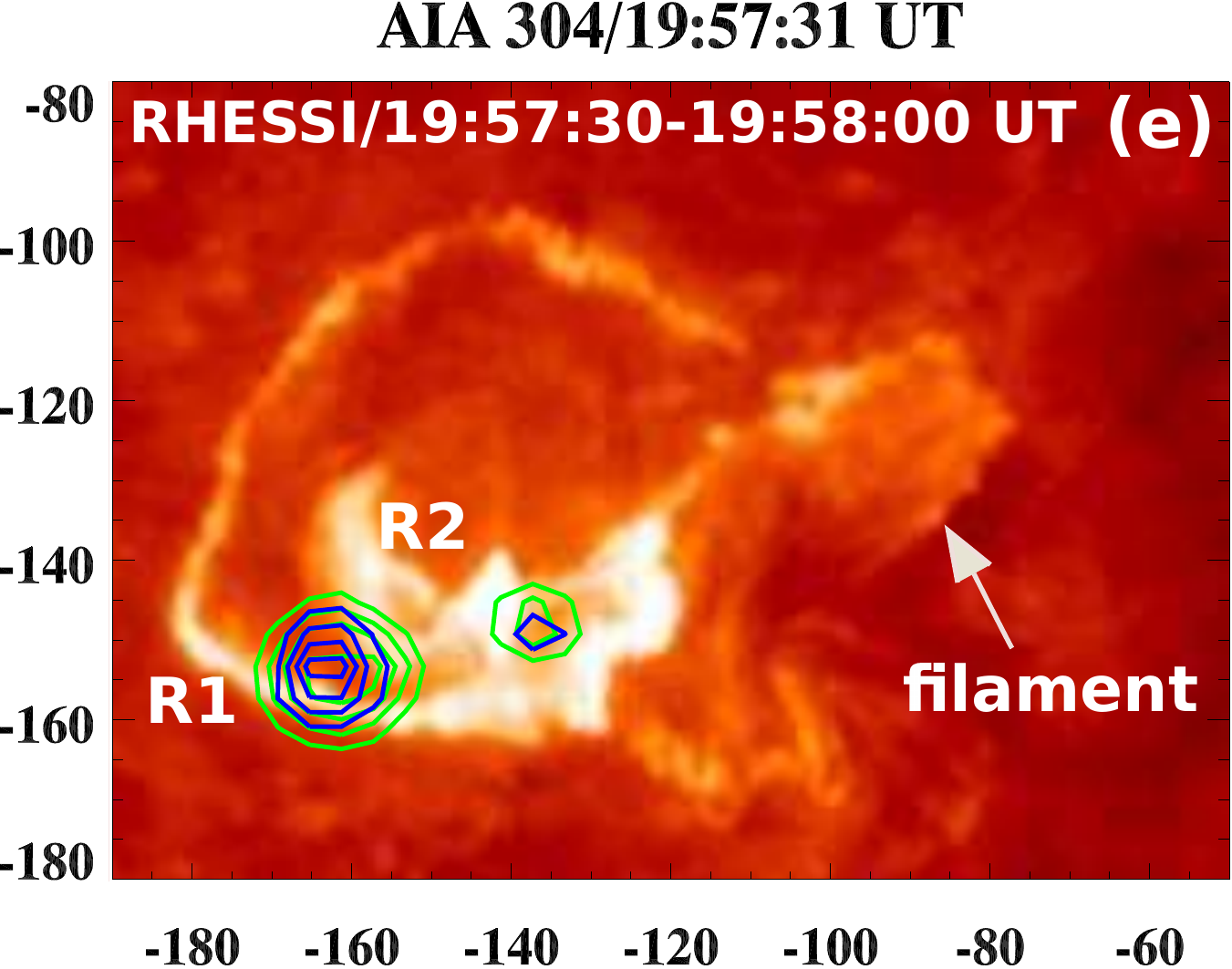}
\includegraphics[width=5.3cm]{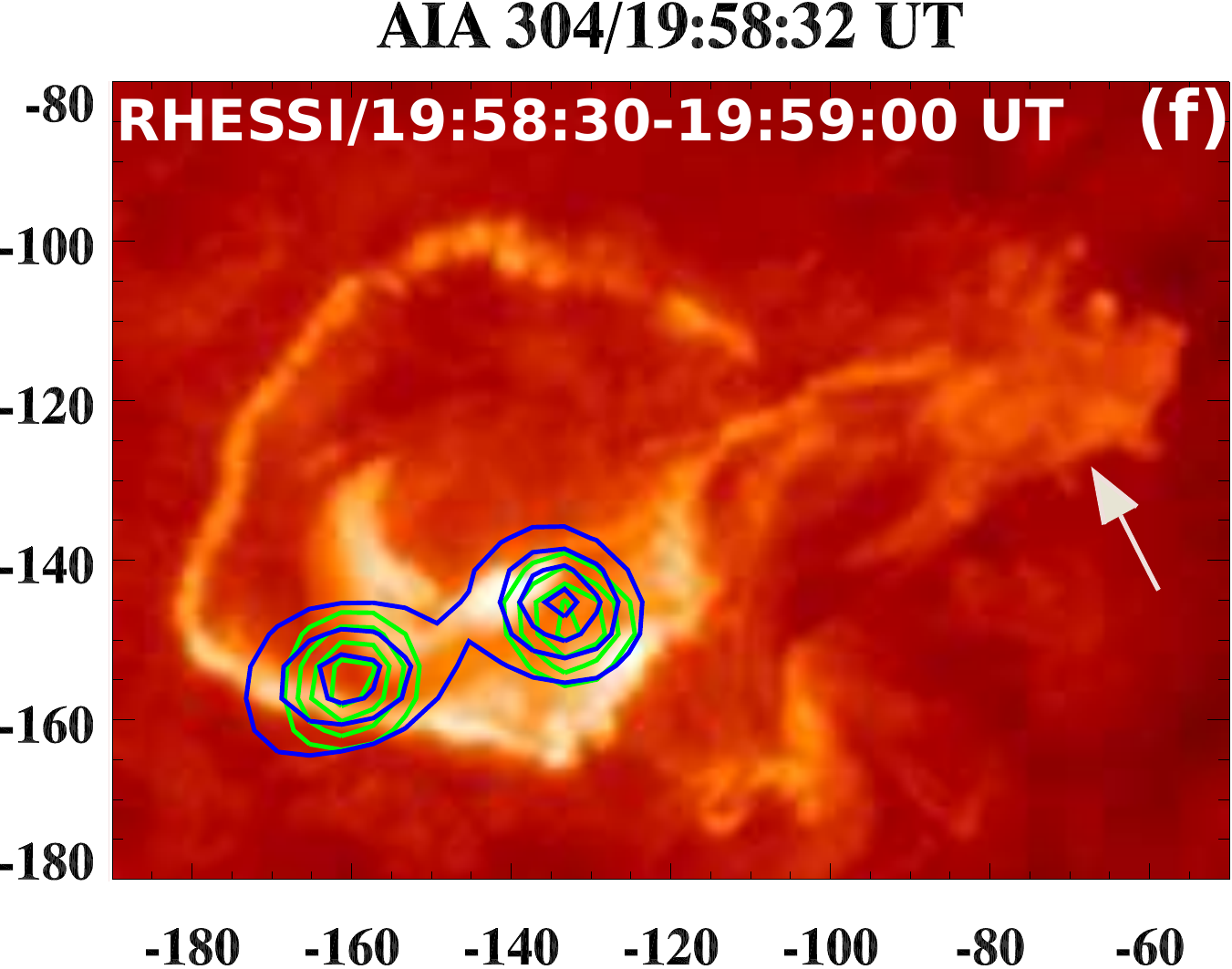}
}
\caption{(a-c) Stack plots of the AIA 94 \AA~ intensity along the slices S5 and S6. The red rectangular box indicates the EIS field of view (refer to the next figure). (d-f) AIA 304 \AA~ images overlaid by RHESSI X-ray  contours in 6-12 keV (green) and 12-25 keV (blue) channels. The contour levels are 30$\%$, 50$\%$, 70$\%$, and 90$\%$ of the peak intensity. HMI magnetogram contours of positive (white) and negative (black) polarities are overlaid in panel (e). R1 and R2 indicate the flare ribbons. (An animation of this figure is available)}
\label{eruption}
\end{figure*}


\begin{figure*}
\centering{
\includegraphics[width=14cm]{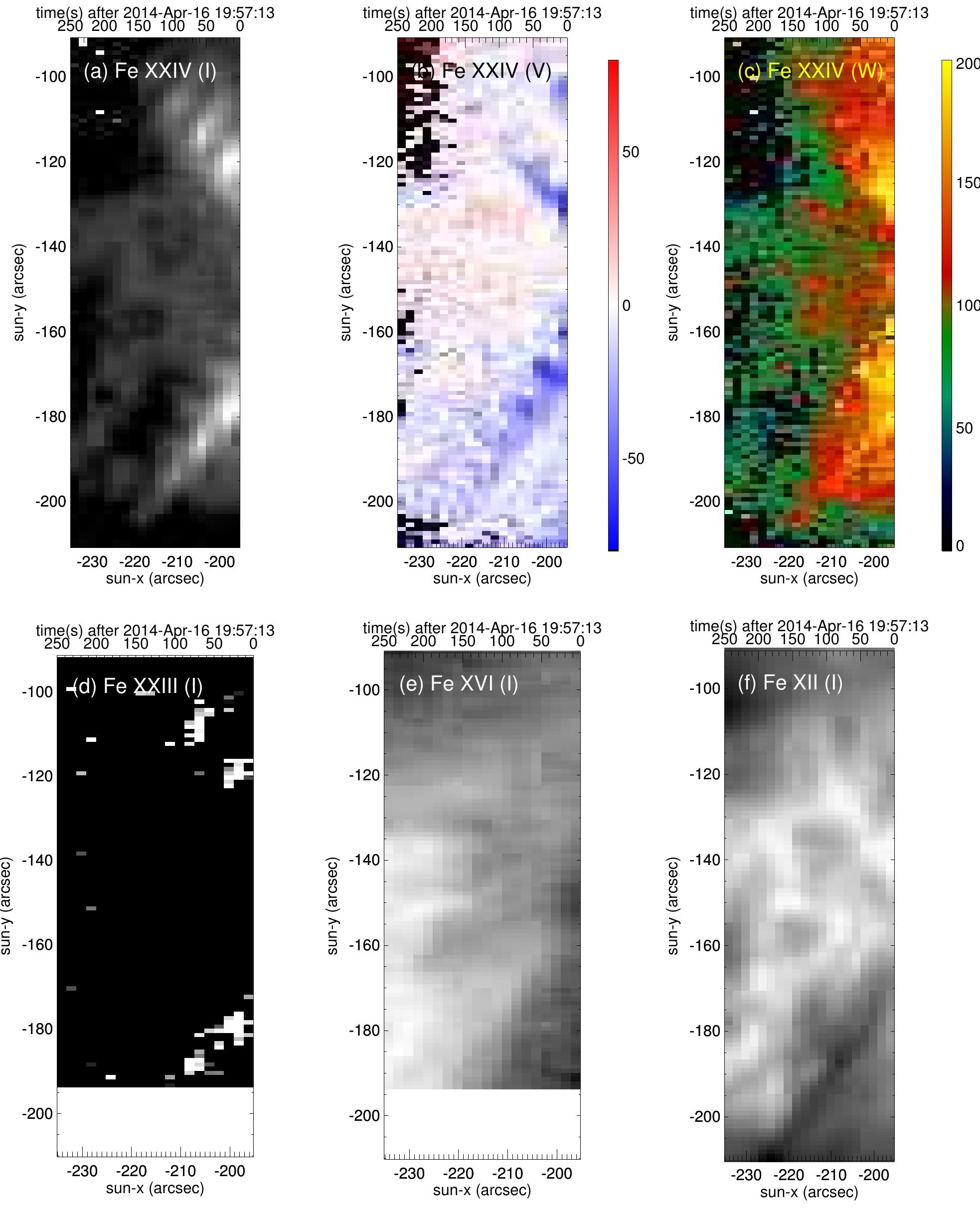}
}
\caption{(a-c) Hinode/EIS \ion{Fe}{24} line intensity, velocity, and width images. (d-f) \ion{Fe}{23}, \ion{Fe}{16}, and \ion{Fe}{12} line intensity images. The start time of the scan is 19:57:13 UT.}
\label{eis}
\end{figure*}

 

\subsection{EUV wave}

Figure \ref{aia-rd} displays running difference images in the AIA 335, 193, and 171~\AA\ channels. The time difference ($\Delta$t) for the difference images is 1 minute. The AIA 335~\AA\ image at 19:56:50 UT shows the EUV disturbance propagating to the left of the flare site. The wavefront is simultaneously observed in the AIA 171 and 193 channels as it propagates through the arcade loops connected to the flare site. Note that the initial phase (from 19:56:38 UT to 19:58:38 UT) of the EUV wavefront was best observed in the AIA 335~\AA\ channel. Later it could not be observed in the AIA 335~\AA\ channel; however, it was observed clearly in the AIA 193 channel from 19:58:42-20:01:35 UT. Figure \ref{aia-rd} (c,d) shows the semi-circular wavefront (F, yellow arrows) moving outward towards the eastern limb after passing through the arcade loops. We did not see a clear wavefront in the AIA 171~\AA\ channel, but a long coronal loop can be seen expanding at 19:56:37~UT and shortly afterwards contracting in the 171~\AA\ images of the AIA 171, 131 and 94  intensity movie  as the loop was crossed by the wave. 
At 19:59:35 UT, we see brightening in the funnel loops (in the AIA 171 channel) possibly caused by compression generated by the propagating wavefront (Figure \ref{aia-rd}(e)) and  later we see a part of the wavefront at 20:00:23 UT (marked by green arrows). The details of the propagating wavefront can be viewed in the AIA 335, 193, and 171 running-difference  movie. The interesting points to be noted here are: (i) the wavefront is highly directive toward the eastern limb, (ii) the wavefront is not moving in all the directions, as generally seen in the case of an expanding CME bubble, (iii) there is only a single wavefront not multiple wave trains, (iv)  transverse loop oscillations are seen throughout the arcade after the passage of the wave through them, (v) the flare occurred at one of the footpoints of the arcade loops, (vi) we do not see any CME loop or flux rope running behind the propagating wavefront in any of the AIA channels. 
 
To estimate the speed of the propagating disturbance, we selected the straight slices S1, S2, and S3 shown in Figure \ref{aia-rd}(a,c) in the AIA 335 and 193 \AA~ channels. The flare center is chosen as the origin for all slices. The time-distance intensity (running difference) plot is displayed in Figure \ref{aia_stack}. S1 and S2 are selected almost in the same direction and the estimated speeds are the projected speeds of the wavefront. From a linear fit, we estimated the speed of the wavefront in the AIA 335 and 193 \AA~ channels, to be $\sim$800$\pm$40 km s$^{-1}$ and $\sim$1490$\pm$190 km s$^{-1}$ respectively. We assumed 5 pixels error in the distance of the wavefront. The speed along slice S3 was $\sim$1360 $\pm$210 km s$^{-1}$. The initial speed of the front in the 193 \AA~ channel (S2) was $\sim$740$\pm$40 km s$^{-1}$. The speed of the semi-circular front (outside the arcade loops) is almost similar in the 193 \AA~ channel and does not change significantly along S2 and S3.

To show the kinematics of the wavefront, we derived the speed of the wavefront using time-distance measurements from the AIA 335 and 193 \AA~ channels (Figure \ref{h-t}(b)). We utilized 3-point Lagrangian interpolation method to estimate the speed of the wavefront. Initially, the wave accelerated ($\sim$700-900 km s$^{-1}$) slowly while passing through the arcade loops. The speed of the wavefront jumped ($\sim$2800 km s$^{-1}$) after leaving the arcade loops and subsequently showed rapid deceleration ($\sim$2800-1100 km s$^{-1}$).

To see the association of the EUV wave with the M1.0 flare, we over-plotted the GOES soft X-ray flux (dashed curve) from the 1-8 \AA~ channel in Figure \ref{aia_stack}(a). The EUV wave onset time ($\sim$19:56 UT)  closely matched the flare impulsive phase. 

The time-distance measurements of the EUV wave in the AIA 335 and 193 \AA~ channels have been over-plotted by the red + symbol and filled circles, respectively in Figure~\ref{h-t}.  The speed of the type II exciter from the Newkirk two-fold model is roughly consistent with the EUV wave speed observed in the AIA 335 \AA~ channel. An exact match cannot be expected because the radio gives the outward speed and the EUV the front speed across the disk. The previous studies have also suggested the consistency of the Newkirk two-fold model with the observed shock wave in the AIA \citep{kumar2013b,kumar2015w}. Therefore, the EUV disturbance (shock wave) passing through the arcade loops is the exciter of the type II radio burst.

\subsection{Transverse oscillations in the arcade loops}
The AIA 335, 193, and 131 \AA~ composite movie (running difference) shows transverse oscillations of many loops within the arcade. These oscillations were triggered when the fast EUV wave passed through the arcade loops. The EUV disturbance seems to be reflected back after  reaching the opposite footpoint of the arcade loops. Although we observed the loop system from above,  it is difficult to say whether it is due to  reflection of the disturbance or transverse oscillations of arcade loops. It is very likely that the fast wave propagates not only along the arcade loops but also across/perpendicular to the loop system \citep{kumar2015w}.

It is quite difficult to extract the oscillation of an individual loop due to the orientation and mixing of the complex loop system in the 193 and 171~\AA\ images. However, some of the loops 
have been heated  and are visible in the 131~\AA\ channel.   We can use one of these to illustrate the oscillations. Figure \ref{osc}(a) shows a stack plot (running difference intensity) along the slice S4 marked on the 131 \AA~ image (Figure \ref{osc}(a)).
In the 131 \AA~ channel, the oscillation starts $\sim$20:03 UT, after the passage of the disturbance through it.  We can clearly see four peaks.

First of all, we determined the positions of the oscillating loop from the stack plot. We subtracted a second order polynomial profile in order to detrend the oscillation profile. We fitted a decaying sine function (y=A. cos[2$\pi$t/T+$\phi$].e$^{-t/\tau}$) to the oscillation profile (red curve). Where, A, $\phi$, T, and $\tau$ are the initial amplitude, phase angle, period, and decay time of the oscillation.
The estimated period and decay time of the oscillation are 210~s and 506~s, respectively.

To determine the phase speed of the wave, we need to measure the loop length. This active region was lying nearly behind the western limb of STEREO-B. Therefore, we need to determine the 3D structure of the loop with a single point observation. We used the curvature radius maximization method \citep{asc2009} to determine the loop length. The estimated loop length was $\sim$0.17 R$_\odot$ (Figure \ref{osc}(b)). The calculated phase speed of the wave (for the fundamental mode) is 2L/P$\approx$1140~km~s$^{-1}$, and the Alfv\'en speed is 840~km~s$^{-1}$.

\subsection{Magnetic configuration}
In the previous sub-sections, we described the wave propagation through the arcade loops toward the eastern limb.  Figure \ref{mf}(a) displays the potential field extrapolation of the active region based on the HMI magnetogram at 19:45 UT (before the flare onset) . The black and white curved lines represent the closed and open field lines, respectively. A small negative polarity region is surrounded by opposite polarity (positive) regions at the flare site, creating a quasi-circular polarity inversion line (PIL). The enlarged view of the flare site (marked by the red box) is shown in Figure \ref{mf}(b). We can see the  fan loops connecting to the negative polarities from the surrounding opposite polarity field region. This morphology is quite similar to the fan-spine topology \citep{pariat2010}.

Figure \ref{mf}(c) shows the Hinode XRT image (Al-poly filter) of the active region before the flare (19:54:12 UT). Two sets of loops are observed here: lower small loops and higher loops, which is consistent with the extrapolated field lines. Note that the higher loops are most affected by the flare and associated EUV wave. Also the fan loops at the flare site are clearly observed in the XRT image (Figure \ref{mf}(d)). These images are overlaid by the HMI magnetogram contours of positive (green) and negative (yellow) polarities to view the connectivity of the coronal loops.

In Figure \ref{mf}(e,f), we display the AIA 1600 \AA~ images during the flare (19:58:16 UT). Interestingly, we see the formation of a quasi-circular ribbon at the flare site. In addition, there is a remote ribbon toward the eastern side of the flare. These ribbons are the precipitation sites of non-thermal electrons accelerated during the flare. 
The formation of a quasi-circular ribbon confirms the fan-spine topology at the flare site \citep{masson2009}. Therefore, reconnection most likely occurred at the null point of the fan-spine topology producing a quasi-circular ribbon. One of the footpoints of the higher loops was connected to the flare site, therefore, accelerated particles propagated along these loops and precipitated to the opposite footpoint to form the remote ribbon.

\subsection{Eruption}
There was a small eruption at the time of the flare. To investigate the eruption, we analyzed AIA 94 and 304 \AA~ images. 
The evolution of the plasma at the site of the eruption is illustrated in 
Figure~\ref{eruption}. Figure~\ref{eruption} shows the stack plots of the AIA 94~\AA\ intensity along the slices S5 and S6 (marked in Figure \ref{eruption}(c)). 
S5 is selected to show the timing along the quasi-circular ribbon (Figure \ref{eruption}(b)).  The first brightening (ribbon) was at $\sim$19:55:30~UT, followed by plasma ejections.
The AIA 94~\AA\ movie shows a sequential brightening that propagates along the quasi-circular ribbon in the counterclockwise direction with a speed $\sim$80$\pm$10 km s$^{-1}$. 

The slice S5 was selected along the direction of the cool plasma ejection.
 Initially there was a small filament along the quasi-circular polarity inversion line (see AIA 304 and 94 composite movie).  At flare onset (19:55:30~UT), the southern end of the filament, overlaid by hard X-ray contours in Figure~\ref{eruption}(c), brightened at all wavelengths and the quasi-circular ribbon appeared, followed by eruption of the filament. The speed of the plasma ejection (i.e., small filament) was $\sim$340$\pm$40 km s$^{-1}$. After the first ejection, we see a series of minor ejections below the filament, which look like multiple plasma blobs or small plasmoids. The AIA 171, 131, 94  intensity movie reveals the formation of postflare loops below the erupting filament. The eruption did not produce a CME. It looks as though the filament material was stopped by the overlying arcades. Therefore, the filament eruption failed (e.g., \citealt{kumar2011failed}).

In the case of a piston-driven shock, the speed of the shock could be 2-3 times the speed of the driver. The wavefront should be initially located ahead of the driver and can decouple (from the driver) later to propagate freely.
An important thing to note is that in this flare eruption and EUV wave move simultaneously in the opposite directions. Also, the speed of the EUV wave is about 3-4 times larger than the speed of the erupting plasma so it is unlikely that the wave was triggered by the filament eruption. If the wave was driven by the small filament; we should have seen the EUV wavefront ahead of the filament at least for some duration as reported by \citet{kumar2013blob}.  
An alternative  mechanism for non-CME type IIs has been recently proposed by \citet{su2015}. They suggest that the rapid expansion of loops following reconnection may provide a suitable piston to generate  shocks in the lower corona.
In this event, we saw no evidence for rapidly expanding loops that could have driven the shock.


Figure \ref{eruption}(d-f) shows the AIA 304~\AA\ images (chromosphere and transition region) overlaid with the RHESSI X-ray contours in 6-12 keV (green) and 12-25 keV (blue) energy channels (refer to Figure \ref{spectrum}(c) for the flux profiles).  To construct the X-ray images, we used the PIXON algorithm \citep{metcalf1996} with an integration time of 30~s for each image. The evolution can be followed in the AIA 304 and 94~\AA\  composite movie.
The RHESSI 12-25 keV contours (Figure~\ref{eruption}(d)) show two footpoint sources. If we compare them with the RHESSI flux profile in 12-25 keV, they are associated with the flare impulsive phase (mostly non-thermal emission). The erupting filament  and two-ribbons (R1 and R2) underneath are marked in Figure~\ref{eruption}(e). Note that R1 is a part of the quasi-circular ribbon. We observed RHESSI footpoint sources (below the filament) in 6-12 and 12-25 keV from 19:57:30-19:59:00 UT. These sources are almost co-spatial with the R1 and R2 ribbons.

The position of the footpoint sources seem to have changed from the impulsive phase to the decay phase probably due to  the change in the particle precipitation site. We speculate that it may be somehow connected with the sequential brightening observed in the counterclockwise direction which may change the particle precipitation site associated with the rise of the small filament. 

As mentioned before, we observed the formation of a remote ribbon (AIA 1600 \AA) and a circular ribbon. Most of the particles accelerated during the flare were confined along the field lines (higher arcade loops) in the low corona. The electron beam from the reconnection site (null point of the fan-spine topology) precipitates downward into the chromosphere (along the fan loops),  forming  a circular ribbon. In addition, high energy electrons probably follow the higher arcade loops and cause a remote ribbon seen in the 1600~\AA\ images in the eastward direction.
     
In addition, we analysed the available Hinode/EIS spectra of the flare. 
The EIS field-of-view is shown in Figure \ref{eruption}(b) by a rectangular box (red color). It only covered  part of the arcade loops located east of the flare site 
but not the flare ribbons.
 These are the arcade loops through which the fast EUV wave propagated.  
Figure \ref{eis}(a-c) displays \ion{Fe}{24} (192.03 \AA, log T=7.1) line intensity, velocity, and width profiles starting at 19:57:13 UT (from right to left). Note that by that time, the fast EUV wave had already passed through these loops (see Figure \ref{aia-rd}(a-b)), and therefore, we see the after-wave signatures in the spectra.  Interestingly, we see  strong blue shifts ($\sim$50-100 km s$^{-1}$) along the higher loops in the \ion{Fe}{24} line. The same loops are also seen in the \ion{Fe}{23} line (263.76 \AA, log T =7) (Figure \ref{eis}(d)) confirming that these are 10~MK loops. The  \ion{Fe}{24} line is considered the most prominent EIS lines during  large flares  \citep{zanna2005}. The observed blue shifted upflows may be generated by chromospheric evaporation along the higher loops. They coincide with hot arcade loops emanating from the flare site seen in AIA 131/94~\AA\ images. The
\ion{Fe}{16} (251.063~\AA, log T =6.4) intensity image shows the higher coronal parts of the lower arcade loops that are also visible in the Hinode XRT image. The \ion{Fe}{12} (193.5 \AA, log T=6.1) intensity image is more patchy because it is emitted from the cooler lower parts of the loops, closer to their photospheric footpoints. 



\section{DISCUSSION AND CONCLUSION}
We reported on the direct observation of a fast EUV wave propagating through coronal arcade loops and its associated type II radio burst (without a CME). Initially, the wave was best observed in the AIA 335~\AA\ with a speed of $\sim$800 km s$^{-1}$ and later in the AIA 193 and 171~\AA\ channels (1490~km~s$^{-1}$). The fast wave was most likely triggered during breakout reconnection in a fan-spine topology. At the flare onset site, a quasi-circular ribbon was seen just before a small, failed-filament eruption ($\sim$340 km s$^{-1}$) and, in the opposite direction, launching of a fast EUV wave across the arcade loops. 
We rule out the small filament as a driver of the EUV wave because (i) the speed was $\sim$3-4 times smaller than the speed of the fast EUV wave, (ii) it moves in the opposite direction, (iii) we do not see the EUV wavefront ahead of the filament apex as expected in the case of piston driven shock.

The propagation of fast EUV waves is generally affected by active regions and coronal holes. The EUV waves may suffer reflection, trapping, and transmission when they encounter with nearby AR or coronal holes \citep{gopal2009,olmedo2012,kumar2013blob,kumar2015w}.

The speed of the EUV wave was initially lower as it passed through the arcade loops. Later after it had escaped, it speeded up. A similar  jump in shock speed was inferred from the dynamic radio spectrum (type II radio burst). 
The jump probably occurred because the arcade loops have higher density than the corona, so the wave could travel faster in the corona.
A shock wave propagating through the high density loops (at higher heights, $\sim$0.1 Rs) could strengthen due to its lower Alfv\'enic speed \citep{cho2013}. The denser overlying loops around the active region may have a relatively weaker field strength, therefore, lower Alfv\'enic speed  \citep{poh2008}. 
This scenario is probably similar to the excitation of a type II radio burst by a shock wave passing through a denser coronal streamer (i.e., lower Alfv\'enic speed regions) \citep{cho2011,kong2012}.
The shock wave is stronger while passing through the denser loops and can excite the fragmented type II radio burst as suggested by \citet{poh2008}. In our case, the type II radio burst structure is fragmented during the passage of fast EUV wave through the denser overlying loops. The wave showed acceleration while passing through the arcade loops as reported in numerical simulations \citep{poh2008}.   

The lower loops (in the AIA 171 \AA~ channel) in the active region did not show transverse oscillation and were almost unaffected by the EUV wave. The higher loops were mainly affected by the EUV wave and exhibit strong transverse oscillation. These higher loops are the most likely candidate for exciting the type II radio burst (second harmonic$>$300 MHz). 

Fast EUV waves generally follow the path of lower Alfv\'enic speed \citep{vrsnak2008} and type II radio bursts could be excited by the flare generated fast shock wave passing through the high density loops at the periphery of the active region. This could be a reason why the EUV wave does not propagate in all directions from the flare center and followed only the denser loops.

 
 
In addition, we did not observe any slow wavefront behind the observed fast EUV wave. The slow wavefronts generally have speeds three times smaller than the fast wave and are usually interpreted as  CME stretched loops (i.e., pseudo wave) running behind the fast wave \citep{chen2002,chen2011,kumar2013b}. Alternatively, a pseudo wave may be explained by the current shell model \citep{delanee2008} and the reconnection front model \citep{attrill2007}. 
In our case, the wave excitation closely matches the flare impulsive phase. This observation reveals that a low coronal type II radio burst can be generated by an impulsive flare ignited shock wave (i.e., blast wave). We agree that most of the low coronal type II radio bursts are generated by a CME. However, our finding suggests that not all the low coronal type II radio bursts may be generated by a CME (piston-driven).
 
As the fast  EUV wave propagated through the coronal arcade loops, it triggered kink oscillations of the loops. This suggests the wave was highly directional.
 According to \citet{hudson2004}, a high directionality is a common characteristic of flare blast waves that trigger kink oscillations of certain loop-system whereas other nearby loops remain stationary \citep{smith1971,warmuth2004}. For one of the loops the kink oscillation had a period of 210~s and a phase speed of $\sim$1140~km~s$^{-1}$. 

For the first time, we see a positive slowly  drifting burst ($\sim$220 MHz) simultaneously with the onset of a type II radio burst and closely associated with magnetic reconnection. The emission source drifts downward with a speed of $\sim$100-130~km~s$^{-1}$. 
 In the AIA 94~\AA\ channel, we observed a co-temporal brightening propagating counterclockwise along the circular ribbon with a projected speed of $\sim$80 km s$^{-1}$. It is likely that the propagating brightening could be evidence of the rotation/slippage of fan loops around the spine. The speed of the radio source is roughly consistent with the rotating brightening. Therefore, we suggest that the downward moving radio source could be generated by the slippage of the field lines that may cause the shift in the reconnection point (inclined motion) resulting in particle acceleration at increasing densities along the slipping field lines. However, more observational studies are needed to validate this interpretation. According to  3D numerical simulation, torsional spine reconnection or slipping reconnection are expected in a fan-spine magnetic configuration \citep{pontin2007,priest2009,masson2009}.
Alternatively, the narrow band positive drifting structure may also be interpreted by a particle acceleration (downward) at a reconnection outflow generated termination shock \citep{aurass2002,mann2009,chen2015} as it was observed simultaneously during the type II radio burst. \citet{mann2009} explained radio emission features (200-400 MHz) as evidence of particle acceleration at the termination shock generated by reconnection outflow.

We also observed a two-ribbon flare within the  circular flare ribbon. This result is consistent with our previous finding of a two-ribbon flare 
(within a global circular ribbon) generated by untwisting small jets produced during the coalescence of two sheared J-shaped H$\alpha$ loops \citep{kumar2015nst}. The magnetic field configuration was quite similar in both cases, i.e., a fan-spine topology with a quasi-circular ribbon \citep{pontin2007,priest2009,masson2009,pariat2010,wang2012}. Here we observed breakout reconnection followed by the failed eruption of a small filament with a speed of $\sim$340 km s$^{-1}$.

In summary, breakout reconnection in a fan-spine topology launched 
a fast-mode MHD shock that propagated perpendicular to the arcade loops generating a type II radio burst. We speculate that a specific magnetic configuration may be an important candidate for the flare ignited shock wave. However, future studies with high resolution observation will shed more light on this issue.
\acknowledgments
We thank the referee for the positive and constructive comments/suggestions that improved the manuscript considerably.
SDO is a mission for NASA's Living With a Star (LWS) program. The SDO data were (partly) provided by the Korean Data Center (KDC) for SDO in cooperation with NASA and SDO/HMI team. RHESSI is a NASA Small Explorer. Hinode is a Japanese mission developed and launched by ISAS/JAXA, with NAOJ as domestic partner and NASA and UKSA as international partners. It is operated by these agencies in co-operation with ESA and NSC (Norway). This work was supported by the ``Operation of Korea Space Weather Center" of KASI and the KASI basic research funds.

\bibliographystyle{apj}
\bibliography{reference}
\end{document}